\documentclass[aps,prx,twocolumn,superscriptaddress, nofootinbib,  longbibliography]{revtex4-2}
\usepackage{graphicx} 
\usepackage{dcolumn}  
\usepackage{bm}    
\usepackage{amssymb}  
\usepackage{amsfonts, amsmath, amssymb, mathtools}
\usepackage{amsmath}
\usepackage{soul}

\usepackage{lipsum}
\usepackage[normalem]{ulem}
\usepackage{commath}
\usepackage{xcolor}
\usepackage{comment}
\usepackage[toc,page]{appendix}
\usepackage{booktabs}
\usepackage[normalem]{ulem}

\usepackage[binary-units=true]{siunitx}
\usepackage{hyperref}
\usepackage{cleveref}
\RequirePackage{textcase}

\begin{document}

\raggedbottom

\newcommand{\yg}[1]{\textcolor{olive}{#1}}
\newcommand{\blue}[1]{\textcolor{black}{#1}}
\newcommand{\markit}[1]{\textcolor{blue}{#1}}
\newcommand{\todo}[1]{\textcolor{red}{#1}}

\title{Direct estimation of arbitrary observables of an oscillator}

\author{Tanjung Krisnanda}
\thanks{Contributed equally to this article.\\
Corresponding author:
{tanjung@nus.edu.sg}}
\author{Fernando Valadares}
\thanks{Contributed equally to this article.}
\affiliation{Centre for Quantum Technologies, National University of Singapore, Singapore}
\author{Kyle Timothy Ng Chu}
\affiliation{Centre for Quantum Technologies, National University of Singapore, Singapore}
\affiliation{Horizon Quantum Computing, Singapore}
\author{Pengtao Song}
\author{Adrian Copetudo}
\author{Clara Yun Fontaine}
\affiliation{Centre for Quantum Technologies, National University of Singapore, Singapore}
\author{Luk\'{a}\v{s} Lachman}
\author{Radim Filip}
\affiliation{Department of Optics, Palack\'{y} University, 17. listopadu 1192/12, 77146 Olomouc, Czech Republic}
\author{Yvonne Y. Gao}
\email[Corresponding author: ]{yvonne.gao@nus.edu.sg}
\affiliation{Centre for Quantum Technologies, National University of Singapore, Singapore}
\affiliation{Department of Physics, National University of Singapore, Singapore}


\begin{abstract}
Quantum harmonic oscillators serve as fundamental building blocks for quantum information processing, particularly in the context of the bosonic circuit quantum electrodynamics (cQED) platform. Conventional methods for extracting oscillator properties rely on predefined analytical gate sequences to access a restricted set of observables or resource-intensive tomography processes. Here, we introduce the Optimized Routine for Estimation of any Observable (OREO), a numerically optimized protocol that maps the expectation value of arbitrary oscillator observables onto that of an ancillary qubit. We demonstrate OREO in a bosonic cQED system as a means to efficiently measure phase-space quadratures and their higher moments, directly obtain faithful non-Gaussianity ranks, and effectively achieve state preparation independent of initial conditions in the oscillator. These results position OREO as a valuable tool for direct and efficient information extraction from bosonic quantum states, unlocking new possibilities for measurement, control, and state preparation in continuous-variable quantum information processing.
\end{abstract}

\maketitle

Quantum harmonic oscillators offer a powerful platform for generating and manipulating photons, which are fundamental resources for a wide range of quantum information processing applications~\cite{braunstein2005quantum}. Among various physical implementations, the bosonic cQED platform, which leverages superconducting microwave cavities coupled to nonlinear ancillary transmon qubit, stands out for its excellent quantum coherence and control versatility~\cite{joshi2021quantum,copetudo_shaping_2024,valadares2026flux}. Thanks to these features, bosonic cQED systems have enabled many remarkable achievements, from the creation of robust logical qubits~\cite{ofek2016extending, sivak2023real, ni2023beating} to the demonstrations of effective quantum metrology~\cite{pan2025realization, deng2024quantum}. 

In typical bosonic cQED devices, information about the harmonic oscillator is extracted via a controlled unitary operation that maps a target property of the oscillator to a distinct transmon state. Three main analytical gate sequences have been developed to enable direct experimental access to the expectation values of the excitation number~\cite{schuster2007resolving}, parity~\cite{sun2014tracking}, and displacement~\cite{campagne2020quantum} operators. Each of these affords a path towards full tomography~\cite{krisnanda2025demonstrating,copetudo2026direct,krisnanda2025experimental,pan2023protecting} \blue{by incorporating many displacement points in phase space}, which then allows experimentalists to extract any target information from the reconstructed oscillator state via post-processing. However, such tomographic approaches are typically resource intensive, and often excessive when the desired information is associated with a specific observable. Thus, it is desirable to explore a more versatile experimental procedure capable of directly estimating arbitrary observables.

\begin{figure}
\centering
\includegraphics{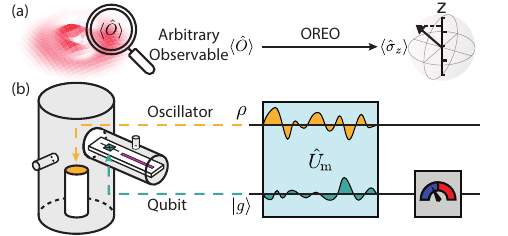}
\caption{{\bf Observable mapping.} 
(a) The Optimize Routine for Estimation of any Observable (OREO) maps the expected value of any target observable $\langle \hat{O}\rangle$ of a harmonic oscillator state onto the z-axis of a qubit's Bloch vector $\langle\hat{\sigma}_z\rangle$. (b) The cQED setup and protocol, which consists of a numerically optimized pulse $\hat{U}_{\text{m}}$ applied to the oscillator-qubit system, followed by a qubit readout. A step-by-step implementation guide is included in the Supplemental Material~\cite{sm}.}
\label{fig1}
\end{figure}

\begin{figure*}[!t]
\centering
\includegraphics{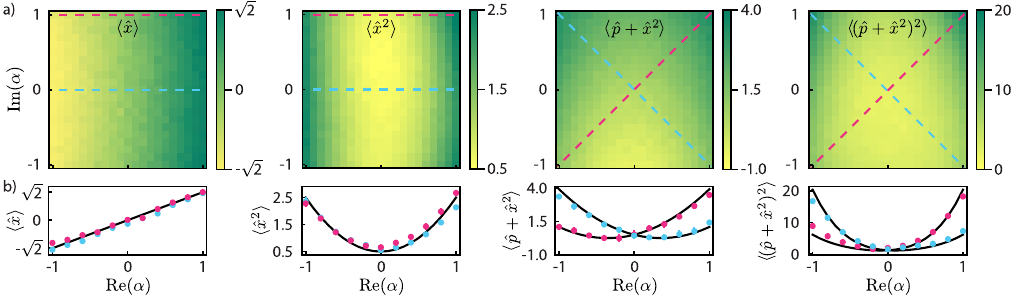}
\caption{{\bf Direct estimation of polynomials of phase-space quadratures.} a) Application of OREO to directly obtain the values of $\langle \hat{x} \rangle$, $\langle \hat{x}^2 \rangle$, $\langle \hat{p}+\hat{x}^2 \rangle$, and $\langle (\hat{p}+\hat{x}^2)^2 \rangle$ of a variable coherent state $|\alpha\rangle$. The dashed lines indicate one-dimensional cuts in the data plotted in (b), which quantitatively agree with the theoretical expected values (solid curves).}
\label{fig2}
\end{figure*}

In this work, we present a one-fits-all technique to estimate the value of any observable associated with a harmonic oscillator via standard qubit measurements. This method, dubbed the Optimized Routine for Estimation of any Observable (OREO), employs numerically optimized pulses to encode the expectation value of a target observable into the state of an ancillary transmon, allowing oscillator properties to be extracted through standard qubit readout (Fig.~\ref{fig1}). OREO is implementable on any oscillator-qubit system that allows universal control on the oscillator space~\cite{sm}, such as trapped ions/atoms~\cite{meekhof1996generation,mccormick2019quantum}, cavity-QED~\cite{hacker2019deterministic}, and quantum acoustics~\cite{wollack2022quantum}.
Compared to the standard Hadamard test, OREO offers a notable reduction in the resources required to obtain the expectation value of an arbitrary Hermitian operator~\cite{sm}.
Here, we experimentally execute OREO in a bosonic cQED architecture, enabling efficient measurement of several useful observables that are typically inaccessible directly. In the first experiment, we probe the phase space quadratures and their higher-order moments, bridging a common gap in bosonic cQED systems, where direct quadrature measurement is typically hindered by the stationary nature of the oscillator mode. In the second, we apply OREO to quantify degrees of non-Gaussianity, providing an efficient means to assess the quality of quantum resources. The last experiment expands on our technique by using the qubit to herald the projection of the oscillator onto an arbitrary state, which affords a strategy to perform robust state preparation independent of the initial condition of the oscillator. 

Our implementation of OREO employs a standard bosonic cQED device~\cite{blais2004cavity}, shown in Fig.~\ref{fig1}, in which a three-dimensional superconducting cavity couples to a transmon qubit in the dispersive regime~\cite{sm}. A planar resonator patterned on the same transmon chip allows direct single-shot measurement of the transmon state. In such a system, universal control of the oscillator state has been demonstrated using numerically optimized drives on both the cavity and transmon~\cite{krastanov2015universal, heeres2017implementing}. Here, we expand the capabilities of this approach to realize direct mapping of the expectation value of any observable of the bosonic mode to the transmon state. 

Conceptually, any observable associated with a quantum harmonic oscillator can be written as $\hat{ \mathcal{O}} = \sum_k \Lambda_k |\psi_k\rangle \langle \psi_k|$, where $\Lambda_k \in \mathbb{R}$ is any real number, and $\{ |\psi_k\rangle \}$ is any basis. Since the qubit measurement restricts the outcomes to be $\langle \hat \sigma_z \rangle \in [-1,1]$, we define a scaled version of the observable, $\hat{O} =\hat{ \mathcal{O}}/f = \sum_k \lambda_k |\psi_k\rangle \langle \psi_k|$ with $f=\max_k |\Lambda_k|$ such that $\lambda_k \in [-1,1]$. 
Next, we numerically optimize a pulse sequence to enact a unitary operation, $\hat U_{\text{m}}(\hat O)$, which is applied to the initial state $|g\rangle \langle g| \otimes \rho$, where $|g\rangle$ is the ground state of the qubit and $\rho$ is an arbitrary state of the oscillator, to implement the mapping process. More specifically, \blue{our framework optimizes the drives to minimize the element-wise mean square error between $\hat O$ and $\hat O_{\text{eff}}$, where \blue{$\hat O_{\text{eff}} \equiv 2\langle g|\hat U_{\text{m}}^{\dagger}|g\rangle \langle g| \hat U_{\text{m}}|g\rangle -\mathbb{I}_{\text{c}}$} denotes an effective observable with $\mathbb{I}_{\text{c}}$ the identity operator within a chosen truncation dimension $D$ where the oscillator state is fully contained~\cite{sm}.} 
Finally, a standard readout is performed on the qubit after $\hat U_{\text{m}}$ to obtain 
\begin{equation}
    \langle \hat \sigma_z \rangle =2p_g-1=\text{tr}_{\text{c}}(\hat O_{\text{eff}} \:\rho), \\
\end{equation}
where $\hat \sigma_z$ denotes the Pauli-$z$ operator, $p_g$ the probability of the qubit in its ground state, $\text{tr}_{\text{c}}$ the trace with respect to the oscillator. This completes the full OREO process and leads to a direct reading of an abitrary $\langle \hat{O}\rangle$~\cite{sm}.
\blue{OREO enables a significantly more efficient estimation of $\langle \hat{O} \rangle$ compared to calculating it from a reconstructed density matrix obtained via full-state tomography.}

To demonstrate the efficacy of OREO in a practical use case, we apply the procedure to directly obtain the expectation values of the phase-space quadratures of the oscillator states residing in the superconducting cavity, as well as their higher moments. Phase-space quadratures are directly accessed in optical systems through homodyne detection and are frequently featured in continuous-variable information processing protocols, such as quantum teleportation, cryptography, and error correction~\cite{braunstein2005quantum}. 
\blue{In contrast, the oscillator states encoded in high-Q cavities are strongly isolated from the environment, a feature that affords excellent lifetimes but hinders direct access to its quadrature information via homodyne detection.}
OREO bridges this gap, providing an efficient and practical method to extract the values of quadrature moments of arbitrary order without special hardware requirements. Access to such information provides a valuable primitive for tasks such as the construction of entanglement witnesses~\cite{mihaescu_detecting_2020}, implementations of nonlinear phase states~\cite{kudra2022robust,eriksson2024universal}, and universal quantum computation via nonlinear squeezing~\cite{moore_estimation_2019, konno2021nonlinear, sakaguchi2023nonlinear}. 

Experimentally, we obtain $\langle \hat \sigma_z\rangle$ from standard single-shot readout after applying drives obtained numerically to map the quadratures and their higher moments onto the qubit. We then perform a simple correction to account for readout errors~\cite{sm}. 
The results for the observables $\langle \hat{x}\rangle$, $\langle \hat{x}^2\rangle $, $\langle \hat{p} + \hat{x}^2\rangle $ and $\langle (\hat{p} + \hat{x}^2)^2\rangle$ for a coherent state $|\alpha\rangle$ displaced to various locations in phase space are shown in Fig.~\ref{fig2}. The data show a strong quantitative agreement with the theoretical values, which shows that OREO is capable of reliably measuring arbitrary quadrature functions. The minor deviations observed for larger displacements, for example $|\alpha|\approx\sqrt{2}$, is due to the leakage of the Fock state population outside the truncation dimension of $D=6$ used in this experiment and can be improved by increasing the truncation dimension~\cite{sm}. 

Next, we apply OREO to benchmark the quality of bosonic quantum states by probing an observable \blue{directly for the first time} to infer their non-Gaussianity rank~\cite{filip2011detecting,lachman_faithful_2019,chabaud_stellar_2020,fiurasek_efficient_2022}, \blue{beyond simply Wigner negativity~\cite{leibfried1996experimental, lutterbach1997method, lvovsky2001quantum, ourjoumtsev2006generating, deleglise2008reconstruction, vlastakis2013deterministically}}. For Fock states and their superpositions, an accurate observable is introduced in Ref.~\cite{fiurasek_efficient_2022}:
\begin{equation}
\label{eq:rankbound}
    \exists \lambda :\langle\hat{\Pi}_n + \lambda\hat{\Pi}_{n+1} \rangle > F_n(\lambda),
\end{equation}
where $F_n(\lambda)$ is a calculated threshold~\cite{sm}, $\hat{\Pi}_n$ is the projector onto the \textit{n}-th Fock state, and $\lambda$ is a free parameter. 
\blue{A state satisfying Eq.~(\ref{eq:rankbound}) means that it could not have been prepared by applying any Gaussian operation on Fock state superpositions involving up to $|n-1\rangle$~\cite{lachman_faithful_2019}. This imposes a hierarchy, certifying the quality of Fock states with higher $n$, which are crucial resources, e.g., for quantum metrology~\cite{wolf2019motional, podhora2022quantum}.}

\begin{figure}
\centering
\includegraphics{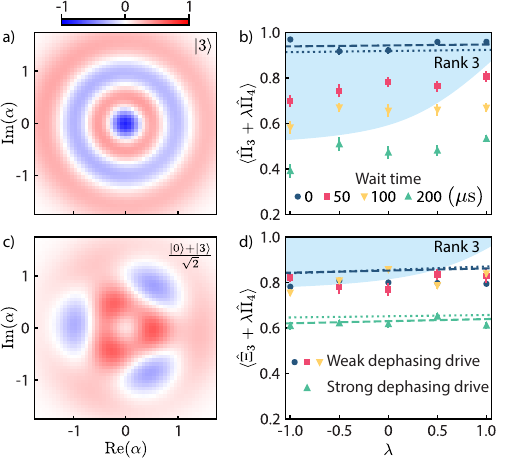}
\caption{{\bf Non-Gaussianity rank.} (a) Reconstructed Wigner of Fock state $|3\rangle$. (b) Measurement of $\langle\hat{\Pi}_3 + \lambda\hat{\Pi}_4 \rangle$ ($\hat{\Pi}_n \equiv |n\rangle\langle n|$) with OREO is used to infer the rank under photon loss for a variable wait time and parameter $\lambda$. The theoretical threshold is shown as the lower boundary of the blue shaded region, above which one infers rank $3$ non-Gaussianity. (c) Reconstructed Wigner for $(|0\rangle + |3\rangle)/\sqrt{2}$, prepared with strong induced dephasing drive. (d) Measurement of $\langle \hat{\Xi}_3 + \lambda \hat{\Pi}_4\rangle$ ($\hat{\Xi}_n \equiv |n\rangle\langle0| + |0\rangle\langle n|$) with OREO under both strong and weak dephasing drives. Both in panels (b) and (d), the outcomes obtained via OREO agree with values derived from the full density matrix reconstruction (dotted lines) \blue{as well as full simulations with decoherence (dashed lines)}.  
}
\label{fig3}
\end{figure}

We investigate this experimentally for two specific non-Gaussian resource states. In the first case, a Fock state $|3\rangle$ was prepared using a standard numerically optimized state transfer process~\cite{heeres2017implementing}. We also performed a standard Wigner tomography and obtained the reconstructed density matrix via Bayesian inference~\cite{krisnanda2025demonstrating}, from which we confirm that a state $|3\rangle$ was prepared with a fidelity of $F=0.92(3)$ (Fig.~\ref{fig3}a). We then perform OREO to directly extract $\langle\hat{\Pi}_3 + \lambda\hat{\Pi}_4 \rangle$, instead of constructing it by separately measuring $\langle \hat \Pi_3\rangle$ and $\langle \hat \Pi_4\rangle$, from single-shot qubit measurements for $\lambda = -1,\,-0.5,\,0,\,0.5,\,1$. The measurement outcomes, shown as blue circles in Fig.~\ref{fig3}b are considerably above the threshold $F_3(\lambda)$ (lower boundary of the blue shaded region), indicating that the state has non-Gaussianity rank of 3. The results, obtained from a single measurement at each value of $\lambda$, are consistent with those derived from the reconstruction of the full density matrix via Wigner tomography (dotted line), while reducing the raw measurement time by nearly three orders of magnitude~\cite{sm}.

To further illustrate the reliability of OREO in identifying the non-Gausianity rank, we execute the same experiment with a variable wait time between state preparation and rank inference, allowing the oscillator state to decay under single photon loss. For a 200-$\mu$s wait time, which is approximately 20\% of single-photon lifetime $T_{1,c}$ of the cavity~\cite{sm}, the measurement outcomes, shown as green triangles, show that $\langle\hat{\Pi}_3 + \lambda\hat{\Pi}_{4} \rangle$ is consistently under the threshold and indicate that we can no longer infer rank 3 non-Gaussianity. For shorter wait times, e.g. 50-$\mu$s and 100-$\mu$s, there are still values of $\lambda$ that satisfy the condition in Eq.~(\ref{eq:rankbound}), signifying that the state still has rank 3. Thus, the flexibility afforded by OREO to extract information on the more complex quantity $\langle\hat{\Pi}_3 + \lambda\hat{\Pi}_4 \rangle$ provides an accurate and efficient mechanism to determine the non-Gaussianity rank and its depth respective to dominant cavity-damping in the system~\cite{straka2014quantum}. 

This ability to reliably extract the non-Gaussianity rank can also reveal subtle imperfections in the resource state that are otherwise not obvious. For instance, we investigate a target state of $(|0\rangle + | 3\rangle)/\sqrt{2}$ in the oscillator and use \blue{a novel} observable $\langle \Xi_3 + \lambda \Pi_4\rangle$ with $\Xi_3 \equiv |3\rangle\langle0| + |0\rangle\langle3|$, which captures the coherence of the quantum superposition, to infer its non-Gaussian coherence rank~\cite{kovalenko2025quantum, asenbeck2025hierarchical, lachman2025hierarchies}. Experimentally, we prepare the target state in the oscillator using a numerically optimized state-transfer pulse~\cite{heeres2017implementing} and then induce a dephasing mechanism on the cavity via a strong pulse on the readout resonator, corresponding to the oscillator's coherence time of $T^{\text{strong}}_{\phi,c}\approx42$~$\mu$s~\cite{sm}.
The reconstructed Wigner \blue{function} of state under strong dephasing drive is shown in Fig.~\ref{fig3}c with fidelity $0.79(2)$ to the ideal target state. It shows three well-defined negative regions that may lead to a naive conclusion that the state has rank 3. \blue{In contrast, when we use OREO to directly probe $\langle \Xi_3 + \lambda \Pi_4\rangle$, the outcome, shown in Fig.~\ref{fig3}d, indicates that the state with strong dephasing drive is consistently below the threshold. This is consistent with the knowledge that strong dephasing has been induced on the cavity state, which should reduce the corresponding non-Gaussianity coherence rank}. Furthermore, as we reduce the strength of the dephasing drive, which results in an improved coherence time of $T^{\text{weak}}_{\phi,c}\approx263$~$\mu$s~\cite{sm}, the values of $\langle \Xi_3 + \lambda \Pi_4\rangle$ \blue{obtained via OREO increase in accordance} and approach the threshold for rank 3. We include data from different iterations of this measurement to verify the consistency and accuracy of the process. \blue{Although} several of these points marginally cross the calculated threshold, we do not have statistically significant evidence to conclude its coherence rank with certainty. Nonetheless, this example highlights the ability of OREO as a sensitive diagnosis tool to assess the coherence of non-Gaussian states. 

\begin{figure}[!h]
\centering
\includegraphics{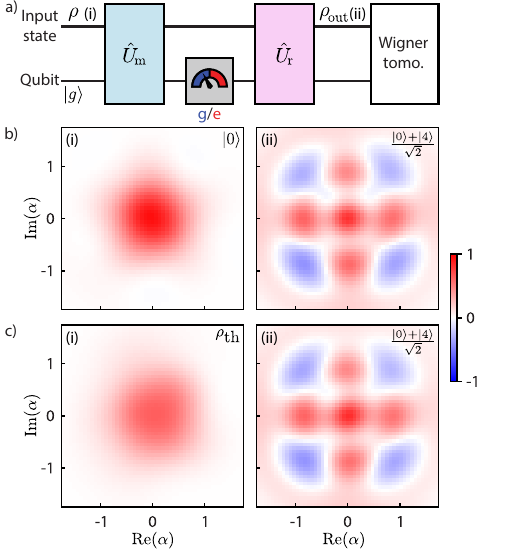}
\caption{
{\bf Projection to a target oscillator state.} a) The protocol that extends OREO, which involves a second numerically optimized pulse $\hat{U}_{\text{r}}$ that reverts the mapping done by $\hat{U}_{\text{m}}$. The resulting oscillator states after successful projections are characterized using Wigner tomography. (b)-(c) The reconstructed Wigner of the initial and final states of the cavity after the projection into the state $|\psi\rangle =(|0\rangle + |4\rangle)/\sqrt{2}$ with the cavity prepared in $|0\rangle$ (b) and in a thermal state with $n_{\text{th}} = 0.24$ (c). The projected oscillator states in (b) and (c) have similar fidelities of $0.84(3)$ and $0.85(2)$, respectively, to the ideal binomial state and non-Gaussianity coherence rank $4$~\cite{sm} despite the differences in initial conditions.}
\label{fig4}
\end{figure}

Finally, we show that OREO can be readily extended to perform a projective measurement and collapse the oscillator state $\rho$ into an arbitrary target $|\psi \rangle$ (Fig.~\ref{fig4}a). \blue{This goes beyond the Fock-state projection, the only currently available technique in cQED in the strong dispersive regime~\cite{schuster2007resolving}.} It builds on the original OREO procedure \blue{with a unitary $\hat U_{\text{m}}$, optimized for estimating the observable $\langle|\psi\rangle \langle \psi|\rangle$, and an additional unitary $\hat{U}_{\text{r}}$ that reverts the mapping of $\hat{U}_{\text{m}}$, optimized such that $\langle g|\hat U_{\text{r}}|g\rangle = \langle g|\hat U_{\text{m}}^{\dagger}|g\rangle $~\cite{sm}.} Here, the intermediate qubit measurement extracts the overlap observable $p=\langle \psi|\rho |\psi  \rangle$ and heralds the correct final state in the oscillator~\cite{sm}. 

As an illustration, we optimize $\hat{U}_{\text{m}}$ and $\hat{U}_{\text{r}}$ for the observable $\hat{\Pi}_\psi = |\psi\rangle\langle\psi|$, where $|\psi\rangle$ is the binomial state $(|0\rangle + |4\rangle)/\sqrt{2}$. 
Measurement of the qubit ground state signals a successful projection to the target state, whereas the oscillator state associated with the qubit in $|e\rangle$ is discarded. We execute this procedure with the oscillator initialized in the vacuum state (Fig.~\ref{fig4}b(i)) and Wigner tomography performed in the end to obtain the reconstructed density matrix of the output state $\rho_{\text{out}}$. The reconstructed Wigner (Fig.~\ref{fig4}b(ii)) shows close agreement with the ideal binomial state, with a success rate of $p=0.49$ and fidelity of $0.84(3)$, limited by qubit Ramsey time and measurement imperfections~\cite{sm}.

Next, we prepare the oscillator in an thermal state (Fig.~\ref{fig4}c(i)), with mean photon number $n_{\text{th}} \approx 0.24$~\cite{sm} with fidelity $0.97(2)$ to the ideal state, and perform the same projective operation as described in Fig.~\ref{fig4}a. Conceptually, a projective operation brings the system to the target state regardless of the initial condition. Here, we observe that the reconstructed Wigner of the output state (Fig.~\ref{fig4}c(ii)) shows a fidelity of $0.85(2)$ to the ideal binomial state, a clear indication that the quality of the projection is not affected by the oscillator's initial condition. However, it comes at the cost of a minor reduction in the success rate ($p=0.44$), which scales with the overlap between the initial and target state. This simple demonstration shows that OREO can be easily extended to implement a projective operation, which serves as an effective oscillator state initialization tool. Although it is a heralded non-deterministic process, its robustness against the initial condition of the oscillator is a valuable feature that is beyond the reach of the standard state-transfer protocols. This is particularly useful for platforms where the oscillator may not have a well-defined initial state due to the presence of residual thermal excitations such as bosonic cQED~\cite{yang2025hot}, opto-mechanical~\cite{aspelmeyer2014cavity}, and quantum acoustic~\cite{wollack2022quantum} devices as well as state generation protocols requiring projective measurements~\cite{lund2004conditional,vasconcelos2010all,eaton2019non,etesse2015experimental,konno2024logical}. 


In conclusion, this work demonstrates the robust capabilities of the Optimized Routine for Estimation of any Observable (OREO) in bridging critical measurement gaps in bosonic cQED systems. By harnessing numerically optimized pulses to encode the expectation value of target observables into an auxiliary transmon qubit, OREO enables single-shot detection of a diverse range of oscillator properties, from phase-space quadratures and their higher-order moments to non-Gaussianity metrics. The performance of OREO is mainly limited by qubit decoherence during the optimized drives, which can potentially be alleviated by known methods to extend the qubit coherence properties~\cite{place2021new,somoroff2023millisecond} or reduce the drive durations~\cite{kudra2022robust}. Even with our modest system parameters~\cite{sm}, experimental results presented in this work \blue{offer compelling validation to} OREO's value and practical \blue{efficacy}. Collectively, these findings underscore OREO as a versatile and effective tool for advancing continuous-variable quantum information processing.

\section*{acknowledgments}
We acknowledge the funding support from the Ministry of Education Singapore (2EP50125-0014 and A-8004168-00-00). F.V., K.C., A.C., and C.F. acknowledge the support of the Singapore National Quantum Scholarship Scheme (NQSS). L.L. acknowledges the support from the project No.~23-06015O and R.F. from the project 21-13265X, both of the Czech Science Foundation. 

\bibliography{Bibliography} 

@misc{sm,
    key = {},
    note = {See the Supplemental Material for details regarding: device and system parameters; proof of universality; implementation of OREO; comparison between OREO and the Hadamard test; simulation of imperfections; calibration and correction for readout assignment error; non-Gaussianity inference; impact of induced dephasing on superpositions of Fock states; preparation of thermal state; sequential projections; and how to use OREO in your experiment. This includes Refs.~\cite{valadares2024demand, blais2021circuit,  sears2012photon, eickbusch2022fast}.}
}

@article{krisnanda2025experimental,
  title={Experimental demonstration of enhanced quantum tomography via quantum reservoir processing},
  author={Krisnanda, Tanjung and Song, Pengtao and Copetudo, Adrian and Fontaine, Clara Yun and Paterek, Tomasz and Liew, Timothy CH and Gao, Yvonne Y},
  journal={Quantum Science and Technology},
  volume={10},
  number={3},
  pages={035041},
  year={2025},
  publisher={IOP Publishing}
}

@article{eickbusch2022fast,
  title={Fast universal control of an oscillator with weak dispersive coupling to a qubit},
  author={Eickbusch, Alec and Sivak, Volodymyr and Ding, Andy Z and Elder, Salvatore S and Jha, Shantanu R and Venkatraman, Jayameenakshi and Royer, Baptiste and Girvin, Steven M and Schoelkopf, Robert J and Devoret, Michel H},
  journal={Nature Physics},
  volume={18},
  number={12},
  pages={1464--1469},
  year={2022},
  publisher={Nature Publishing Group UK London}
}

@article{sun2014tracking,
  title={Tracking photon jumps with repeated quantum non-demolition parity measurements},
  author={Sun, Luyan and Petrenko, Andrei and Leghtas, Zaki and Vlastakis, Brian and Kirchmair, Gerhard and Sliwa, KM and Narla, Aniruth and Hatridge, Michael and Shankar, Shyam and Blumoff, Jacob and others},
  journal={Nature},
  volume={511},
  number={7510},
  pages={444--448},
  year={2014},
  publisher={Nature Publishing Group UK London}
}

@article{schuster2007resolving,
  title={Resolving photon number states in a superconducting circuit},
  author={Schuster, DI and Houck, Andrew Addison and Schreier, JA and Wallraff, A and Gambetta, JM and Blais, A and Frunzio, L and Majer, J and Johnson, B and Devoret, MH and others},
  journal={Nature},
  volume={445},
  number={7127},
  pages={515--518},
  year={2007},
  publisher={Nature Publishing Group UK London}
}

@article{campagne2020quantum,
  title={Quantum error correction of a qubit encoded in grid states of an oscillator},
  author={Campagne-Ibarcq, Philippe and Eickbusch, Alec and Touzard, Steven and Zalys-Geller, Evan and Frattini, Nicholas E and Sivak, Volodymyr V and Reinhold, Philip and Puri, Shruti and Shankar, Shyam and Schoelkopf, Robert J and others},
  journal={Nature},
  volume={584},
  number={7821},
  pages={368--372},
  year={2020},
  publisher={Nature Publishing Group UK London}
}

@article{krastanov2015universal,
  title={Universal control of an oscillator with dispersive coupling to a qubit},
  author={Krastanov, Stefan and Albert, Victor V and Shen, Chao and Zou, Chang-Ling and Heeres, Reinier W and Vlastakis, Brian and Schoelkopf, Robert J and Jiang, Liang},
  journal={Physical Review A},
  volume={92},
  number={4},
  pages={040303},
  year={2015},
  publisher={APS}
}

@article{heeres2017implementing,
  title={Implementing a universal gate set on a logical qubit encoded in an oscillator},
  author={Heeres, Reinier W and Reinhold, Philip and Ofek, Nissim and Frunzio, Luigi and Jiang, Liang and Devoret, Michel H and Schoelkopf, Robert J},
  journal={Nature communications},
  volume={8},
  number={1},
  pages={94},
  year={2017},
  publisher={Nature Publishing Group UK London}
}

@article{valadares2024demand,
  title={On-demand transposition across light-matter interaction regimes in bosonic cQED},
  author={Valadares, Fernando and Huang, Ni-Ni and Chu, Kyle Timothy Ng and Dorogov, Aleksandr and Chua, Weipin and Kong, Lingda and Song, Pengtao and Gao, Yvonne Y},
  journal={Nature Communications},
  volume={15},
  number={1},
  pages={5816},
  year={2024},
  publisher={Nature Publishing Group UK London}
}

@article{ofek2016extending,
  title={Extending the lifetime of a quantum bit with error correction in superconducting circuits},
  author={Ofek, Nissim and Petrenko, Andrei and Heeres, Reinier and Reinhold, Philip and Leghtas, Zaki and Vlastakis, Brian and Liu, Yehan and Frunzio, Luigi and Girvin, Steven M and Jiang, Liang and others},
  journal={Nature},
  volume={536},
  number={7617},
  pages={441--445},
  year={2016},
  publisher={Nature Publishing Group UK London}
}

@article{sivak2023real,
  title={Real-time quantum error correction beyond break-even},
  author={Sivak, VV and Eickbusch, Alec and Royer, Baptiste and Singh, Shraddha and Tsioutsios, Ioannis and Ganjam, Suhas and Miano, Alessandro and Brock, BL and Ding, AZ and Frunzio, Luigi and others},
  journal={Nature},
  volume={616},
  number={7955},
  pages={50--55},
  year={2023},
  publisher={Nature Publishing Group UK London}
}

@article{ni2023beating,
  title={Beating the break-even point with a discrete-variable-encoded logical qubit},
  author={Ni, Zhongchu and Li, Sai and Deng, Xiaowei and Cai, Yanyan and Zhang, Libo and Wang, Weiting and Yang, Zhen-Biao and Yu, Haifeng and Yan, Fei and Liu, Song and others},
  journal={Nature},
  volume={616},
  number={7955},
  pages={56--60},
  year={2023},
  publisher={Nature Publishing Group UK London}
}

@article{chabaud_stellar_2020,
  title={Stellar representation of non-Gaussian quantum states},
  author={Chabaud, Ulysse and Markham, Damian and Grosshans, Fr{\'e}d{\'e}ric},
  journal={Physical Review Letters},
  volume={124},
  number={6},
  pages={063605},
  year={2020},
  publisher={APS}
}

@article{fiurasek_efficient_2022,
  title={Efficient construction of witnesses of the stellar rank of nonclassical states of light},
  author={Fiur{\'a}{\v{s}}ek, Jarom{\'\i}r},
  journal={Optics Express},
  volume={30},
  number={17},
  pages={30630--30639},
  year={2022},
  publisher={Optica Publishing Group}
}

@article{copetudo_shaping_2024,
  title={Shaping photons: Quantum information processing with bosonic cQED},
  author={Copetudo, Adrian and Fontaine, Clara Yun and Valadares, Fernando and Gao, Yvonne Y},
  journal={Applied Physics Letters},
  volume={124},
  pages={080502},
  number={8},
  year={2024},
  publisher={AIP Publishing}
}

@article{mihaescu_detecting_2020,
  title={Detecting entanglement of unknown continuous variable states with random measurements},
  author={Mihaescu, Tatiana and Kampermann, Hermann and Gianfelici, Giulio and Isar, Aurelian and Bru{\ss}, Dagmar},
  journal={New Journal of Physics},
  volume={22},
  number={12},
  pages={123041},
  year={2020},
  publisher={IOP Publishing}
}

@article{lachman_faithful_2019,
  title={Faithful hierarchy of genuine n-photon quantum non-Gaussian light},
  author={Lachman, Luk{\'a}{\v{s}} and Straka, Ivo and Hlou{\v{s}}ek, Josef and Je{\v{z}}ek, Miroslav and Filip, Radim},
  journal={Physical review letters},
  volume={123},
  number={4},
  pages={043601},
  year={2019},
  publisher={APS}
}

@article{moore_estimation_2019,
  title={Estimation of squeezing in a nonlinear quadrature of a mechanical oscillator},
  author={Moore, Darren W and Rakhubovsky, Andrey A and Filip, Radim},
  journal={New Journal of Physics},
  volume={21},
  number={11},
  pages={113050},
  year={2019},
  publisher={IOP Publishing}
}

@article{konno2021nonlinear,
  title={Nonlinear squeezing for measurement-based non-gaussian operations in time domain},
  author={Konno, Shunya and Sakaguchi, Atsushi and Asavanant, Warit and Ogawa, Hisashi and Kobayashi, Masaya and Marek, Petr and Filip, Radim and Yoshikawa, Jun-ichi and Furusawa, Akira},
  journal={Physical Review Applied},
  volume={15},
  number={2},
  pages={024024},
  year={2021},
  publisher={APS}
}

@article{sakaguchi2023nonlinear,
  title={Nonlinear feedforward enabling quantum computation},
  author={Sakaguchi, Atsushi and Konno, Shunya and Hanamura, Fumiya and Asavanant, Warit and Takase, Kan and Ogawa, Hisashi and Marek, Petr and Filip, Radim and Yoshikawa, Jun-ichi and Huntington, Elanor and others},
  journal={Nature Communications},
  volume={14},
  number={1},
  pages={3817},
  year={2023},
  publisher={Nature Publishing Group UK London}
}

@article{pan2025realization,
  title={Realization of versatile and effective quantum metrology using a single bosonic mode},
  author={Pan, Xiaozhou and Krisnanda, Tanjung and Duina, Andrea and Park, Kimin and Song, Pengtao and Fontaine, Clara Yun and Copetudo, Adrian and Filip, Radim and Gao, Yvonne Y},
  journal={PRX Quantum},
  volume={6},
  number={1},
  pages={010304},
  year={2025},
  publisher={APS}
}

@article{deng2024quantum,
  title={Quantum-enhanced metrology with large Fock states},
  author={Deng, Xiaowei and Li, Sai and Chen, Zi-Jie and Ni, Zhongchu and Cai, Yanyan and Mai, Jiasheng and Zhang, Libo and Zheng, Pan and Yu, Haifeng and Zou, Chang-Ling and others},
  journal={Nature Physics},
  volume={20},
  pages={pages1874},
  year={2024},
  publisher={Nature Publishing Group UK London}
}

@article{joshi2021quantum,
  title={Quantum information processing with bosonic qubits in circuit QED},
  author={Joshi, Atharv and Noh, Kyungjoo and Gao, Yvonne Y},
  journal={Quantum Science and Technology},
  volume={6},
  number={3},
  pages={033001},
  year={2021},
  publisher={IOP Publishing}
}

@article{braunstein2005quantum,
  title={Quantum information with continuous variables},
  author={Braunstein, Samuel L and Van Loock, Peter},
  journal={Reviews of modern physics},
  volume={77},
  number={2},
  pages={513--577},
  year={2005},
  publisher={APS}
}

@article{krisnanda2025demonstrating,
  title={Demonstrating efficient and robust bosonic state reconstruction via optimized excitation counting},
  author={Krisnanda, Tanjung and Fontaine, Clara Yun and Copetudo, Adrian and Song, Pengtao and Lee, Kai Xiang and Huang, Ni-Ni and Valadares, Fernando and Liew, Timothy CH and Gao, Yvonne Y},
  journal={PRX Quantum},
  volume={6},
  number={1},
  pages={010303},
  year={2025},
  publisher={APS}
}

@article{pan2023protecting,
  title={Protecting the quantum interference of cat states by phase-space compression},
  author={Pan, Xiaozhou and Schwinger, Jonathan and Huang, Ni-Ni and Song, Pengtao and Chua, Weipin and Hanamura, Fumiya and Joshi, Atharv and Valadares, Fernando and Filip, Radim and Gao, Yvonne Y},
  journal={Physical Review X},
  volume={13},
  number={2},
  pages={021004},
  year={2023},
  publisher={APS}
}

@article{blais2004cavity,
  title={Cavity quantum electrodynamics for superconducting electrical circuits: An architecture for quantum computation},
  author={Blais, Alexandre and Huang, Ren-Shou and Wallraff, Andreas and Girvin, Steven M and Schoelkopf, R Jun},
  journal={Physical Review A—Atomic, Molecular, and Optical Physics},
  volume={69},
  number={6},
  pages={062320},
  year={2004},
  publisher={APS}
}

@article{wollack2022quantum,
  title={Quantum state preparation and tomography of entangled mechanical resonators},
  author={Wollack, E Alex and Cleland, Agnetta Y and Gruenke, Rachel G and Wang, Zhaoyou and Arrangoiz-Arriola, Patricio and Safavi-Naeini, Amir H},
  journal={Nature},
  volume={604},
  number={7906},
  pages={463--467},
  year={2022},
  publisher={Nature Publishing Group UK London}
}

@article{aspelmeyer2014cavity,
  title={Cavity optomechanics},
  author={Aspelmeyer, Markus and Kippenberg, Tobias J and Marquardt, Florian},
  journal={Reviews of Modern Physics},
  volume={86},
  number={4},
  pages={1391--1452},
  year={2014},
  publisher={APS}
}

@article{kudra2022robust,
  title={Robust preparation of Wigner-negative states with optimized SNAP-displacement sequences},
  author={Kudra, Marina and Kervinen, Mikael and Strandberg, Ingrid and Ahmed, Shahnawaz and Scigliuzzo, Marco and Osman, Amr and Lozano, Daniel P{\'e}rez and Thol{\'e}n, Mats O and Borgani, Riccardo and Haviland, David B and others},
  journal={PRX Quantum},
  volume={3},
  number={3},
  pages={030301},
  year={2022},
  publisher={APS}
}

@article{place2021new,
  title={New material platform for superconducting transmon qubits with coherence times exceeding 0.3 milliseconds},
  author={Place, Alexander PM and Rodgers, Lila VH and Mundada, Pranav and Smitham, Basil M and Fitzpatrick, Mattias and Leng, Zhaoqi and Premkumar, Anjali and Bryon, Jacob and Vrajitoarea, Andrei and Sussman, Sara and others},
  journal={Nature communications},
  volume={12},
  number={1},
  pages={1779},
  year={2021},
  publisher={Nature Publishing Group UK London}
}

@article{somoroff2023millisecond,
  title={Millisecond coherence in a superconducting qubit},
  author={Somoroff, Aaron and Ficheux, Quentin and Mencia, Raymond A and Xiong, Haonan and Kuzmin, Roman and Manucharyan, Vladimir E},
  journal={Physical Review Letters},
  volume={130},
  number={26},
  pages={267001},
  year={2023},
  publisher={APS}
}

@article{sears2012photon,
  title = {Photon shot noise dephasing in the strong-dispersive limit of circuit QED},
  author = {Sears, A. P. and Petrenko, A. and Catelani, G. and Sun, L. and Paik, Hanhee and Kirchmair, G. and Frunzio, L. and Glazman, L. I. and Girvin, S. M. and Schoelkopf, R. J.},
  journal = {Phys. Rev. B},
  volume = {86},
  issue = {18},
  pages = {180504},
  numpages = {5},
  year = {2012},
  month = {Nov},
  publisher = {American Physical Society}
}

@article{blais2021circuit,
  title = {Circuit quantum electrodynamics},
  author = {Blais, Alexandre and Grimsmo, Arne L. and Girvin, S. M. and Wallraff, Andreas},
  journal = {Rev. Mod. Phys.},
  volume = {93},
  issue = {2},
  pages = {025005},
  numpages = {72},
  year = {2021},
  month = {May},
  publisher = {American Physical Society}
}

@article{straka2014quantum,
  title={Quantum non-Gaussian depth of single-photon states},
  author={Straka, Ivo and Predojevi{\'c}, Ana and Huber, Tobias and Lachman, Luk{\'a}{\v{s}} and Butschek, Lorenz and Mikov{\'a}, Martina and Mi{\v{c}}uda, Michal and Solomon, Glenn S and Weihs, Gregor and Je{\v{z}}ek, Miroslav and others},
  journal={Physical Review Letters},
  volume={113},
  number={22},
  pages={223603},
  year={2014},
  publisher={APS}
}

@article{eriksson2024universal,
  title={Universal control of a bosonic mode via drive-activated native cubic interactions},
  author={Eriksson, Axel M and S{\'e}pulcre, Th{\'e}o and Kervinen, Mikael and Hillmann, Timo and Kudra, Marina and Dupouy, Simon and Lu, Yong and Khanahmadi, Maryam and Yang, Jiaying and Castillo-Moreno, Claudia and others},
  journal={Nature Communications},
  volume={15},
  number={1},
  pages={2512},
  year={2024},
  publisher={Nature Publishing Group UK London}
}

@article{filip2011detecting,
  title={Detecting quantum states with a positive Wigner function beyond mixtures of Gaussian states},
  author={Filip, Radim and Mi{\v{s}}ta Jr, Ladislav},
  journal={Physical review letters},
  volume={106},
  number={20},
  pages={200401},
  year={2011},
  publisher={APS}
}

@article{meekhof1996generation,
  title={Generation of nonclassical motional states of a trapped atom},
  author={Meekhof, DM and Monroe, C and King, BE and Itano, Wayne M and Wineland, David J},
  journal={Physical review letters},
  volume={76},
  number={11},
  pages={1796},
  year={1996},
  publisher={APS}
}

@article{mccormick2019quantum,
  title={Quantum-enhanced sensing of a single-ion mechanical oscillator},
  author={McCormick, Katherine C and Keller, Jonas and Burd, Shaun C and Wineland, David J and Wilson, Andrew C and Leibfried, Dietrich},
  journal={Nature},
  volume={572},
  number={7767},
  pages={86--90},
  year={2019},
  publisher={Nature Publishing Group UK London}
}

@article{hacker2019deterministic,
  title={Deterministic creation of entangled atom--light Schr{\"o}dinger-cat states},
  author={Hacker, Bastian and Welte, Stephan and Daiss, Severin and Shaukat, Armin and Ritter, Stephan and Li, Lin and Rempe, Gerhard},
  journal={Nature Photonics},
  volume={13},
  number={2},
  pages={110--115},
  year={2019},
  publisher={Nature Publishing Group UK London}
}

@article{lund2004conditional,
  title={Conditional production of superpositions of coherent states with inefficient photon detection},
  author={Lund, AP and Jeong, H and Ralph, TC and Kim, MS},
  journal={Physical Review A—Atomic, Molecular, and Optical Physics},
  volume={70},
  number={2},
  pages={020101},
  year={2004},
  publisher={APS}
}

@article{vasconcelos2010all,
  title={All-optical generation of states for “Encoding a qubit in an oscillator”},
  author={Vasconcelos, Hilma M and Sanz, Liliana and Glancy, Scott},
  journal={Optics letters},
  volume={35},
  number={19},
  pages={3261--3263},
  year={2010},
  publisher={Optical Society of America}
}

@article{eaton2019non,
  title={Non-Gaussian and Gottesman--Kitaev--Preskill state preparation by photon catalysis},
  author={Eaton, Miller and Nehra, Rajveer and Pfister, Olivier},
  journal={New Journal of Physics},
  volume={21},
  number={11},
  pages={113034},
  year={2019},
  publisher={IOP Publishing}
}

@article{etesse2015experimental,
  title={Experimental generation of squeezed cat states with an operation allowing iterative growth},
  author={Etesse, Jean and Bouillard, Martin and Kanseri, Bhaskar and Tualle-Brouri, Rosa},
  journal={Physical review letters},
  volume={114},
  number={19},
  pages={193602},
  year={2015},
  publisher={APS}
}

@article{konno2024logical,
  title={Logical states for fault-tolerant quantum computation with propagating light},
  author={Konno, Shunya and Asavanant, Warit and Hanamura, Fumiya and Nagayoshi, Hironari and Fukui, Kosuke and Sakaguchi, Atsushi and Ide, Ryuhoh and China, Fumihiro and Yabuno, Masahiro and Miki, Shigehito and others},
  journal={Science},
  volume={383},
  number={6680},
  pages={289--293},
  year={2024},
  publisher={American Association for the Advancement of Science}
}

@article{yang2025hot,
  title={Hot Schr{\"o}dinger cat states},
  author={Yang, Ian and Agrenius, Thomas and Usova, Vasilisa and Romero-Isart, Oriol and Kirchmair, Gerhard},
  journal={Science Advances},
  volume={11},
  number={14},
  pages={eadr4492},
  year={2025},
  publisher={American Association for the Advancement of Science}
}

@article{leibfried1996experimental,
  title={Experimental determination of the motional quantum state of a trapped atom},
  author={Leibfried, Dietrich and Meekhof, DM and King, BE and Monroe, CH and Itano, Wayne M and Wineland, David J},
  journal={Physical Review Letters},
  volume={77},
  number={21},
  pages={4281},
  year={1996},
  publisher={APS}
}

@article{lutterbach1997method,
  title={Method for direct measurement of the Wigner function in cavity QED and ion traps},
  author={Lutterbach, LG and Davidovich, L},
  journal={Physical Review Letters},
  volume={78},
  number={13},
  pages={2547},
  year={1997},
  publisher={APS}
}

@article{lvovsky2001quantum,
  title={Quantum state reconstruction of the single-photon Fock state},
  author={Lvovsky, Alexander I and Hansen, Hauke and Aichele, T and Benson, O and Mlynek, J and Schiller, S},
  journal={Physical Review Letters},
  volume={87},
  number={5},
  pages={050402},
  year={2001},
  publisher={APS}
}

@article{ourjoumtsev2006generating,
  title={Generating optical Schrodinger kittens for quantum information processing},
  author={Ourjoumtsev, Alexei and Tualle-Brouri, Rosa and Laurat, Julien and Grangier, Philippe},
  journal={Science},
  volume={312},
  number={5770},
  pages={83--86},
  year={2006},
  publisher={American Association for the Advancement of Science}
}

@article{deleglise2008reconstruction,
  title={Reconstruction of non-classical cavity field states with snapshots of their decoherence},
  author={Deleglise, Samuel and Dotsenko, Igor and Sayrin, Clement and Bernu, Julien and Brune, Michel and Raimond, Jean-Michel and Haroche, Serge},
  journal={Nature},
  volume={455},
  number={7212},
  pages={510--514},
  year={2008},
  publisher={Nature Publishing Group UK London}
}

@article{vlastakis2013deterministically,
  title={Deterministically encoding quantum information using 100-photon Schr{\"o}dinger cat states},
  author={Vlastakis, Brian and Kirchmair, Gerhard and Leghtas, Zaki and Nigg, Simon E and Frunzio, Luigi and Girvin, Steven M and Mirrahimi, Mazyar and Devoret, Michel H and Schoelkopf, Robert J},
  journal={Science},
  volume={342},
  number={6158},
  pages={607--610},
  year={2013},
  publisher={American Association for the Advancement of Science}
}

@article{wolf2019motional,
  title={Motional Fock states for quantum-enhanced amplitude and phase measurements with trapped ions},
  author={Wolf, Fabian and Shi, Chunyan and Heip, Jan C and Gessner, Manuel and Pezz{\`e}, Luca and Smerzi, Augusto and Schulte, Marius and Hammerer, Klemens and Schmidt, Piet O},
  journal={Nature communications},
  volume={10},
  number={1},
  pages={2929},
  year={2019},
  publisher={Nature Publishing Group UK London}
}

@article{podhora2022quantum,
  title={Quantum non-Gaussianity of multiphonon states of a single atom},
  author={Podhora, Luk{\'a}{\v{s}} and Lachman, Luk{\'a}{\v{s}} and Pham, Tuan and Le{\v{s}}und{\'a}k, Adam and {\v{C}}{\'\i}p, Ond{\v{r}}ej and Slodi{\v{c}}ka, Luk{\'a}{\v{s}} and Filip, Radim},
  journal={Physical Review Letters},
  volume={129},
  number={1},
  pages={013602},
  year={2022},
  publisher={APS}
}

@article{kovalenko2025quantum,
  title={Quantum non-Gaussian coherences of an oscillating atom},
  author={Kovalenko, A and Lachman, L and Pham, T and Singh, K and {\v{C}}{\'\i}p, O and Fiur{\'a}{\v{s}}ek, J and Slodi{\v{c}}ka, L and Filip, R},
  journal={Physical Review Research},
  volume={7},
  number={3},
  pages={033075},
  year={2025},
  publisher={APS}
}

@article{asenbeck2025hierarchical,
  title={Hierarchical Verification of Non-Gaussian Coherence in Bosonic Quantum States},
  author={Asenbeck, Beate E and Lachman, Luk{\'a}{\v{s}} and Boyer, Ambroise and Giri, Priyanka and Urvoy, Alban and Filip, Radim and Laurat, Julien},
  journal={Physical Review Letters},
  volume={134},
  number={23},
  pages={233604},
  year={2025},
  publisher={APS}
}

@article{lachman2025hierarchies,
  title={Hierarchies of quantum non-Gaussian coherences for bosonic systems: A theoretical study},
  author={Lachman, Luk{\'a}{\v{s}} and Asenbeck, Beate E and Boyer, Ambroise and Giri, Priyanka and Urvoy, Alban and Laurat, Julien and Filip, Radim},
  journal={Physical Review A},
  volume={111},
  number={6},
  pages={063704},
  year={2025},
  publisher={APS}
}

@article{valadares2026flux,
  title={Flux-Activated Resonant Control of a Bosonic Quantum Memory},
  author={Valadares, Fernando and Dorogov, Aleksandr and Krisnanda, Tanjung and Loke, May Chee and Huang, Ni-Ni and Song, Pengtao and Gao, Yvonne Y},
  journal={arXiv preprint arXiv:2602.18122},
  year={2026}
}

@article{copetudo2026direct,
  title={A direct controlled-phase gate between microwave photons},
  author={Copetudo, Adrian and Kasper, Amon M and Krisnanda, Tanjung and Veyrac, Gregoire and Qin, Shushen and Ng, Hui Khoon and Gao, Yvonne Y},
  journal={arXiv preprint arXiv:2603.15587},
  year={2026}
}

\clearpage
\newpage

\begin{center}
\bf SUPPLEMENTAL MATERIAL
\end{center}

\section{Device \& System Parameters}
The device used in this work (Fig.~1 in the main text) is a standard bosonic cQED system operating in the strong-dispersive coupling regime. It consists of a superconducting 3D microwave cavity, an ancillary transmon and a planar readout resonator. The cavity was machined from a block of high-purity (4N) aluminum, while the transmon and readout resonator were fabricated by evaporating aluminum on a sapphire substrate.

The Hamiltonian of the system can be written as
\begin{equation}
    \hat H = \hat H_0 + \hat H_{d} \label{EQ_fullHamiltonian},
\end{equation}
with the bare and drive terms respectively expressed as
\begin{eqnarray}
\hat H_0/\hbar &=& \sum_{k=q,c,r} \omega_k \hat k^{\dagger} \hat k-\frac{\chi_{kk}}{2}\hat k^{\dagger} \hat k^{\dagger} \hat k \hat k \nonumber \\
&&-\chi_{qc}  \hat q^{\dagger}\hat q \hat c^{\dagger}\hat c-\chi_{qr}  \hat q^{\dagger}\hat q \hat r^{\dagger}\hat r -\chi_{cr} \hat c^{\dagger}\hat c \hat r^{\dagger}\hat r \nonumber \\
&&-\chi_{qc}^{\prime}  \hat q^{\dagger}\hat q \hat c^{\dagger}\hat c^{\dagger}\hat c \hat c,
\end{eqnarray}
and
\begin{eqnarray}
\hat H_{d}/\hbar &=& \epsilon_{q}(t) \hat qe^{i\omega_{dq}t} + \epsilon_{q}^*(t) \hat q^{\dagger}e^{-i\omega_{dq}t} \nonumber \\
&&+\epsilon_{c}(t) \hat c e^{i\omega_{dc}t} + \epsilon_{c}^*(t) \hat c^{\dagger}e^{-i\omega_{dc}t},
\end{eqnarray}
where we have used $k$ to indicate the qubit ($q$), oscillator ($c$), and resonator ($r$). The operator $\hat k$ ($\hat k^{\dagger}$) denotes the annihilation (creation) operator of the corresponding system with frequency $\omega_k$ and self-Kerr $\chi_{kk}$, $\chi_{qc}$ the qubit-oscillator cross-Kerr or dispersive coupling, $\chi_{qr}$ the qubit-resonator cross-Kerr, $\chi_{cr}$ the oscillator-resonator cross-Kerr, $\chi_{qc}^{\prime}$ the second-order qubit-oscillator dispersive coupling, and $\epsilon_{q}(t)$ ($\epsilon_{c}(t)$) the time-dependent complex drive amplitude of the qubit (oscillator) with frequency $\omega_{dq}$ ($\omega_{dc}$). The experimental values of the Hamiltonian parameters in our system are summarized in Table ~\ref{table_para}.
Note that as the self-Kerr (anharmonicity) of the transmon is high, only the first two energy levels are occupied and it can effectively be treated as a qubit with $|g\rangle$ ($|e\rangle$) as its ground (excited) state.

\begin{table}[!h]
\begin{tabular}{l|c|c}
\hline
\hline
Description                                      & Label & Value \\
\hline
Cavity frequency                                 & $\omega_c/2\pi$ & 4.641\,GHz  \\
\hline
Transmon frequency                               & $\omega_q/2\pi$ & 5.327\,GHz  \\
\hline
Reasonator frequency                             & $\omega_r/2\pi$ & 7.568\,GHz  \\
\hline
Cavity self-Kerr                                 & $\chi_{cc}/2\pi$ & 4.4\,kHz     \\
\hline
Transmon self-Kerr                           & $\chi_{qq}/2\pi$ & 175.6\,MHz \\
\hline
Transmon-cavity cross-Kerr                      & $\chi_{qc}/2\pi$ & 1.482\,MHz  \\
\hline
2nd order correction to $\chi_{qc}$              & $\chi'_{qc}/2\pi$ & 13.6\,kHz   \\
\hline
Transmon-resonator cross-Kerr                   & $\chi_{qr}/2\pi$ & 0.49\,MHz \\
\hline
Cavity $T_1$                                        & $T_{1,c}$ & 1-1.1\,ms  \\
\hline
Transmon $T_1$                                      & $T_{1,q}$ & 100\,$\mu$s   \\
\hline
Transmon $T_2$ Ramsey                            & $T_{2,q}$ & 40$\,\mu$s   \\
\hline
\hline
\end{tabular}
\caption{\textbf{Hamiltonian parameters and coherence times of the bosonic cQED device.}}\label{table_para}
\end{table}

\section{Proof of Universality}
Here we show that any observable $\hat{ \mathcal{O}}$ of a harmonic oscillator in any state $\rho$ can be estimated by applying a unitary operation $\hat U_{\text{m}}$, from the qubit-oscillator Hamiltonian in Eq.~(\ref{EQ_fullHamiltonian}) with optimized drives in time, and a qubit measurement. 

We start by noting that any observable can be written as $\hat{ \mathcal{O}} = \sum_k \Lambda_k |\psi_k\rangle \langle \psi_k|$, where $\Lambda_k \in \mathbb{R}$ is any real number, and $\{ |\psi_k\rangle \}$ is any basis. Since the qubit measurement restricts the outcomes to be $\langle \hat \sigma_z \rangle \in [-1,1]$, we will measure the scaled version
\begin{equation}
    \hat{O} =\hat{ \mathcal{O}}/f = \sum_k \lambda_k |\psi_k\rangle \langle \psi_k|,
\end{equation}
where $f=\max_k |\Lambda_k|$ such that $\lambda_k \in [-1,1]$. After the measurements, we obtain the expectation value of the unscaled observable as $\mbox{tr}(\hat{ \mathcal{O}} \rho)\equiv \langle \hat{ \mathcal{O}} \rangle  = f \times \langle \hat O \rangle$.

Let the initial qubit-oscillator state be \( |g\rangle \langle g| \otimes \rho \), and define the unitary \( \hat{U}_{\text{m}} = \hat{U}_{\text{P}} \hat{U} \), where \( \hat{U}_{\text{P}} \) implements the Ramsey sequence, which, together with the qubit measurement, realizes the standard parity measurement of the oscillator state~\cite{sun2014tracking}. The parity can then be written as
\begin{eqnarray}
    P &=& \text{tr}(\mathbb{I}_2\otimes \hat P \: \hat U \: |g\rangle \langle g|\otimes \rho \:\hat U^{\dagger})\nonumber \\
    &=& \text{tr}(\mathbb{I}_2\otimes \hat P \: |g\rangle \langle g|\otimes \hat U_{\text{c}}\:\rho \:\hat U_{\text{c}}^{\dagger})\nonumber \\
    &=&\text{tr}_{\text{c}}(\hat U_{\text{c}}^{\dagger} \hat P \hat U_{\text{c}} \: \rho) \nonumber \\
    &\equiv&\text{tr}_{\text{c}}(\hat O_{\text{eff}} \: \rho),\label{EQ_obspar}
\end{eqnarray}
where $\hat P\equiv \exp(i\hat c^{\dagger}\hat c \pi)$ denotes the parity operator and $\text{tr}_{\text{c}}$ the trace with respect to the oscillator, and the steps are explained as follows:
for the second line, we use the fact that the Hamiltonian in Eq.~(\ref{EQ_fullHamiltonian}) with optimized drives can realize any unitary $\hat U_{\text{c}}$ on the oscillator~\cite{krastanov2015universal}, i.e., $\hat U \: |g\rangle \langle g|\otimes \rho \:\hat U^{\dagger}=|g\rangle \langle g|\otimes \hat U_{\text{c}}\:\rho \:\hat U_{\text{c}}^{\dagger}$; the cyclic property of trace and partial trace with respect to the qubit are used in the third line; and effective observable is defined in the fourth line as $\hat O_{\text{eff}}=\hat U_{\text{c}}^{\dagger} \hat P \hat U_{\text{c}}$.

In realistic situations, one can choose a sufficient truncation dimension $D$ within which the arbitrary state of the oscillator can be fully expressed. Consequently, the observable of interest is also truncated within $D$, which can be arbitrarily chosen. Now we show that any observable within dimension $D$,
\begin{equation}
    \hat O =\sum_{k=0}^{D-1} \lambda_k |\psi_k\rangle \langle \psi_k| \label{EQ_obsscaledtrun}
\end{equation}
with $\{|\psi_k\rangle \}$ being a basis within $D$, can be achieved by applying the universal unitary on the parity operator, as defined in Eq.~(\ref{EQ_obspar}).
First, we note that the parity operator can be expressed as $\hat P=\sum_{n=0}^{\infty} (-1)^{n}|n\rangle \langle n|$ with $\{|n\rangle \}$ being the Fock basis. As any unitary on the oscillator is possible, there exists $\hat U_{\text{c}}^{\dagger}$ that applies the following transformation
\begin{eqnarray}
    |0\rangle &\rightarrow& 
    \alpha_0 |\psi_0\rangle +\beta_0 |h_1\rangle \nonumber \\
    |h_1\rangle &\rightarrow& 
    \beta_0 |\psi_0\rangle -\alpha_0 |h_1\rangle \nonumber \\
    |1\rangle &\rightarrow& 
    \alpha_1 |\psi_1\rangle +\beta_1 |h_2\rangle \nonumber \\
    |h_2\rangle &\rightarrow& 
    \beta_1 |\psi_1\rangle -\alpha_1 |h_2\rangle \nonumber \\
    &\vdots&\label{EQ_basistrans}
\end{eqnarray}
for any $\alpha_n,\beta_n \in \mathbb{R}$ with $\alpha_n^2+\beta_n^2=1$, where we have used $|h_n\rangle$ to denote Fock state outside of the truncation dimension $D$, starting from $|h_1\rangle$ with parity $-1$. It can be seen that the LHS and RHS in Eqs.~(\ref{EQ_basistrans}) each consists of states that form a complete basis. Within the dimension $D$, we then have 
\begin{equation}
    \hat O_{\text{eff}} = \sum_{k=0}^{D-1} (-1)^k (\alpha_k^2-\beta_k^2) |\psi_k\rangle \langle\psi_k|,
\end{equation}
where $(-1)^k(\alpha_k^2-\beta_k^2) \in [-1,1]$. These coefficients can be arbitrary for all $k$s as any unitary $\hat U_{\text{c}}^{\dagger}$ is possible, which completes the proof. 

Additionally, the relation between parity and qubit measurement, following the Ramsey sequence~\cite{sun2014tracking}, is given by $P=-\langle \hat \sigma_z\rangle=1-2p_g$, where $p_g$ is the probability of the qubit being in the ground state. If one post processes $p_g$ instead of $\langle \hat \sigma_z\rangle$, we have $p_g = (1-P)/2=\text{tr}_{\text{c}}(\sum_k (1-\lambda_k)/2 \:|\psi_k\rangle \langle \psi_k|\: \rho)$, where $(1-\lambda_k)/2 \in [0,1]$ and the corresponding observable $\sum_k (1-\lambda_k)/2 \:|\psi_k\rangle \langle \psi_k|$ is positive.

\section{Implementation of OREO}

Here, we provide the detailed recipe of the OREO protocol for the estimation of any observable, as depicted in Fig.~1, as well as the state projection protocol, as depicted in Fig.~4a. 

\subsection{Observable estimation}

For estimation of an arbitrary observable we apply an optimized unitary $\hat U_{\text{m}}$ to the initial state $|g\rangle \langle g| \otimes \rho$ followed by a qubit measurement. The latter can then be written as
\begin{eqnarray}
    \langle \hat \sigma_z \rangle &=&2p_g-1\nonumber \\
    &=&2\text{tr}(|g\rangle \langle g|\otimes \mathbb{I}_{\text{c}}\:\hat U_{\text{m}}\: |g\rangle \langle g| \otimes \rho \: \hat U_{\text{m}}^{\dagger})-1\nonumber \\
    &=&\text{tr}_{\text{c}}\left((2\langle g|\hat U_{\text{m}}^{\dagger}|g\rangle \langle g| \hat U_{\text{m}}|g\rangle -\mathbb{I}_{\text{c}})\rho\right), 
\end{eqnarray}
where we have used the partial trace with respect to the qubit and the cyclic property of trace.
Given an arbitrary observable $\hat {\mathcal{O}}$ with $\hat O$ as its scaled version truncated to dimension $D$ (see Eq.~(\ref{EQ_obsscaledtrun})), we optimize the unitary $\hat U_{\text{m}}$ such that $\hat O_{\text{eff}}=\hat O$, where 
\begin{equation}
    \hat O_{\text{eff}}\equiv 2\langle g|\hat U_{\text{m}}^{\dagger}|g\rangle \langle g| \hat U_{\text{m}}|g\rangle -\mathbb{I}_{\text{c}} \label{EQ_Oeffsigmaz}
\end{equation} 
truncated to dimension $D$. \blue{Ideally, we then have $\langle \hat \sigma_z \rangle=\text{tr}_{\text{c}}(\hat O_{\text{eff}} \:\rho)$. We note that the equality holds at the level of expectation values. Although single-shot qubit measurements are binary, their statistical average converges to the expectation value of the scaled observable.}

As explained in the previous section, if the observable is positive, one can post process $p_g$ instead of $\langle \hat \sigma_z \rangle$. In this case, we have $p_g=\text{tr}_{\text{c}}(\langle g|\hat U_{\text{m}}^{\dagger}|g\rangle \langle g| \hat U_{\text{m}}|g\rangle \:\rho)$, and the effective observable now reads 
\begin{equation}
    \hat O_{\text{eff}}\equiv \langle g|\hat U_{\text{m}}^{\dagger}|g\rangle \langle g| \hat U_{\text{m}}|g\rangle \label{EQ_Oeffpg}
\end{equation} 
truncated to dimension $D$.

The optimization is performed similar to the GRAPE method~\cite{heeres2017implementing}, i.e., optimizing the amplitude of the drives $\{\epsilon_q(t),\epsilon_c(t)\}$ in time with gradient descent to minimize a cost function.
In the present case, the cost function is taken as the element-wise mean square error
\begin{equation}
    \mathcal{C} = \frac{1}{D^2}\sum_{ij}|(\hat O-\hat O_{\text{eff}})_{ij}|^2,\label{EQ_costfobs}
\end{equation}
where $ij$ denotes the $(i,j)$-th matrix element.
This cost function ensures that all elements of $\hat O_{\text{eff}}$ be close to those of the target observable $\hat O$ within the chosen truncation dimension $D$. 

\subsection{Two-element POVM}
We would like to note that OREO is not equivalent to a general multi-element positive operator-valued measure (POVM), as we are only measuring an ancillary qubit with two outcomes, corresponding to ground and excited states. In this case, one can see that OREO can implement any two-element POVM $\{\hat E_1, \hat E_2\}$ as follows. Leading to Eq.~(\ref{EQ_Oeffpg}) we have
\begin{eqnarray}
    p_g&=&\text{tr}_{\text{c}}(\langle g|\hat U_{\text{m}}^{\dagger}|g\rangle \langle g| \hat U_{\text{m}}|g\rangle \:\rho)\nonumber \\
    &\equiv&\text{tr}_{\text{c}}(\hat E_1 \:\rho),
\end{eqnarray}
where $\hat E_1\equiv \langle g|\hat U_{\text{m}}^{\dagger}|g\rangle \langle g| \hat U_{\text{m}}|g\rangle$ can be any positive operator as shown in the previous section. Similarly, for the qubit in the excited state, we have
\begin{eqnarray}
    p_e&=&\text{tr}_{\text{c}}(\langle g|\hat U_{\text{m}}^{\dagger}|e\rangle \langle e| \hat U_{\text{m}}|g\rangle \:\rho)\nonumber \\
    &\equiv&\text{tr}_{\text{c}}(\hat E_2 \:\rho), 
\end{eqnarray}
where $\hat E_2 \equiv \langle g|\hat U_{\text{m}}^{\dagger}|e\rangle \langle e| \hat U_{\text{m}}|g\rangle$. It is easy to see that $\hat E_1+\hat E_2=\mathbb{I}_{\text{c}}$.

\subsection{State projection}

Here, we will focus on the case where $\hat O=|\psi\rangle \langle\psi|$. This corresponds to a projective measurement, where the goal is twofold: to extract the probability of overlap $\langle \psi|\rho |\psi\rangle$ and have the oscillator state collapse to $|\psi \rangle$. The sequence for this is presented in Fig.~4a, where a second unitary $\hat{U}_r$ is introduced to obtain the collapsed cavity state.

As the observable is positive, to get the probability of overlap, we use Eq.~(\ref{EQ_Oeffpg}), where the qubit measurement reads
\begin{eqnarray}
    p_g &=&\text{tr}_{\text{c}}(\langle g|\hat U_{\text{m}}^{\dagger}|g\rangle \langle g| \hat U_{\text{m}}|g\rangle \:\rho) \nonumber \\
    &=&\text{tr}_{\text{c}}(|\psi\rangle \langle \psi| \:\rho) \nonumber \\
    &=& \langle \psi| \rho |\psi\rangle.
\end{eqnarray}

In a single-shot manner, state of the oscillator after the qubit measurement $\tilde \rho_{\text{c}|j}$ depends on the outcome, whether the qubit is in the ground state $j=g$ or excited state $j=e$. In the case of the former, the qubit-oscillator state after the qubit measurement can be written as
\begin{eqnarray}
    \tilde \rho_{\text{qc}|g} &=&\frac{|g\rangle \langle g| \otimes \mathbb{I}_{\text{c}}\:\hat U_{\text{m}}\:|g\rangle \langle g| \otimes \rho \: \hat U_{\text{m}}^{\dagger}\:|g\rangle \langle g| \otimes \mathbb{I}_{\text{c}}}{p_g}\nonumber \\
    &=&|g\rangle \langle g| \otimes \frac{\langle g|\hat U_{\text{m}}|g\rangle \rho \langle g| \hat U_{\text{m}}^{\dagger}|g\rangle}{p_g}\nonumber \\
    &=&|g\rangle \langle g| \otimes \tilde \rho_{\text{c}|g}.
\end{eqnarray}
We note that although the oscillator state at this point is pure, as one can confirm that $\text{tr}(\tilde \rho_{\text{c}|g}^2)=1$, it is not yet in the target projected state $|\psi \rangle$. This is where the second unitary $\hat U_{\text{r}}$ comes into play, after which the qubit-oscillator state can be written as
\begin{widetext}
\begin{eqnarray}
    \rho_{\text{qc}|g} &=&\frac{\hat U_{\text{r}}\:|g\rangle \langle g| \otimes \mathbb{I}_{\text{c}}\:\hat U_{\text{m}}\:|g\rangle \langle g| \otimes \rho \: \hat U_{\text{m}}^{\dagger}\:|g\rangle \langle g| \otimes \mathbb{I}_{\text{c}}\:\hat U_{\text{r}}^{\dagger}}{p_g}.
\end{eqnarray}
\end{widetext}
The probability of the qubit being in ground state here is 
\begin{eqnarray}
    p_{g|g} &=& \text{tr}(|g\rangle \langle g| \otimes \mathbb{I}_{\text{c}}\: \rho_{\text{qc}|g})
    \nonumber \\
    &=&\text{tr}_{\text{c}}\left(\frac{\langle g|\hat U_{\text{r}}|g\rangle \langle g|\hat U_{\text{m}}|g\rangle \rho\langle g|\hat U_{\text{m}}^{\dagger}|g\rangle \langle g|\hat U_{\text{r}}^{\dagger}|g\rangle}{p_g}\right),
\end{eqnarray}
where we carried out the partial trace with respect to the qubit. 
If the second unitary can be optimized such that $\hat U_{\text{r}}=\hat U_{\text{m}}^{\dagger}$, then the probability reduces to
\begin{eqnarray}
    p_{g|g}&=&\text{tr}_{\text{c}}\left(\frac{|\psi\rangle \langle\psi| \rho|\psi\rangle \langle\psi|}{p_g}\right)\nonumber \\
    &=&1,
\end{eqnarray}
which means the qubit is in its ground state decoupled from the oscillator, i.e., $\rho_{\text{qc}|g}=|g\rangle\langle g|\otimes \rho_{\text{out}|g}$.
It is easy to see that 
\begin{eqnarray}
    \rho_{\text{out}|g}&=&\langle g|\rho_{\text{qc}|g}|g\rangle \nonumber \\
    &=&|\psi \rangle \langle \psi|.
\end{eqnarray}
To arrive at this collapsed state, it is not necessary to have $\hat U_{\text{r}}=\hat U_{\text{m}}^{\dagger}$. In fact, it is sufficient to optimize the reverse unitary in the oscillator space only such that $\langle g|\hat U_{\text{r}}|g\rangle=\langle g|\hat U_{\text{m}}^{\dagger}|g\rangle$, which is what we implemented in our experiments. 
In particular, given the unitary $\hat U_{\text{m}}$ that is optimized previously, we optimize the reverse unitary in the same way as described in Eq.~(\ref{EQ_costfobs}) by minimizing the cost function 
\begin{equation}
    \mathcal{C} = \frac{1}{D^2}\sum_{ij}|(\hat M_1-\hat M_{2})_{ij}|^2,\label{EQ_costfunitary},
\end{equation}
where $\hat M_1$ $(\hat M_2)$ denotes $\langle g|\hat U_{\text{m}}^{\dagger}|g\rangle$ ($\langle g|\hat U_{\text{r}}|g\rangle$) truncated to dimension $D$.

In the case where the qubit measurement yields excited state, the qubit-oscillator state after the operation $\hat U_{\text{r}}$ reads
\begin{widetext}
\begin{eqnarray}
    \rho_{\text{qc}|e} &=&\frac{\hat U_{\text{r}}\:|e\rangle \langle e| \otimes \mathbb{I}_{\text{c}}\:\hat U_{\text{m}}\:|g\rangle \langle g| \otimes \rho \: \hat U_{\text{m}}^{\dagger}\:|e\rangle \langle e| \otimes \mathbb{I}_{\text{c}}\:\hat U_{\text{r}}^{\dagger}}{1-p_g}.
\end{eqnarray}
\end{widetext}
The probability of the qubit being in the ground state here is 
\begin{eqnarray}
    p_{g|e} &=& \text{tr}(|g\rangle \langle g| \otimes \mathbb{I}_{\text{c}}\: \rho_{\text{qc}|e})
    \nonumber \\
    &=&\text{tr}_{\text{c}}\left(\frac{\langle g|\hat U_{\text{r}}|e\rangle \langle e|\hat U_{\text{m}}|g\rangle \rho\langle g|\hat U_{\text{m}}^{\dagger}|e\rangle \langle e|\hat U_{\text{r}}^{\dagger}|g\rangle}{1-p_g}\right),
\end{eqnarray}
where the partial trace with respect to the qubit has been carried out. If $\hat U_{\text{r}}=\hat U_{\text{m}}^{\dagger}$, the probability reduces to
\begin{eqnarray}
    p_{g|e}&=&\text{tr}_{\text{c}}\left(\frac{(\mathbb{I}_{\text{c}}-|\psi\rangle \langle\psi|) \rho(\mathbb{I}_{\text{c}}-|\psi\rangle \langle\psi|)}{1-p_g}\right)\nonumber \\
    &=&1,
\end{eqnarray}
where we have used $\langle g|\hat U_{\text{m}}^{\dagger}|e\rangle \langle e|\hat U_{\text{m}}|g\rangle=\langle g|\hat U_{\text{m}}^{\dagger}(\mathbb{I}_2-|g\rangle \langle g|)\hat U_{\text{m}}|g\rangle=\mathbb{I}_c-|\psi\rangle \langle \psi|$. Consequently, the qubit-oscillator state is of the form $\rho_{\text{qc}|e}=|g\rangle \langle g| \otimes \rho_{\text{out}|e}$, where 
\begin{eqnarray}
    \rho_{\text{out}|e}&=&\langle g|\rho_{\text{qc}|e}|g\rangle \nonumber \\
    &=&\frac{(\mathbb{I}_{\text{c}}-|\psi\rangle \langle\psi|) \rho(\mathbb{I}_{\text{c}}-|\psi\rangle \langle\psi|)}{1-p_g}\nonumber \\
    &=&\frac{1}{1-\langle \psi|\rho |\psi \rangle} \sum_{|i\rangle,|j\rangle \ne |\psi\rangle} \langle i|\rho|j\rangle \:|i\rangle \langle j|.\label{EQ_notpsi}
\end{eqnarray}
This is the \emph{not}-$|\psi \rangle$ state, which erases the $|\psi \rangle$ state component from $\rho$ and retains all elements orthogonal to $|\psi \rangle$. For example, if the input states we are dealing with are qubit codes $\{|\psi\rangle,|\phi\rangle\}$, i.e., two orthogonal states encoded in the oscillator (popular for bosonic error correction schemes~\cite{ofek2016extending,sivak2023real,ni2023beating}), the outcome of the qubit in excited state will correspond to the state $|\phi \rangle$. However, equation (\ref{EQ_notpsi}) requires the complete inversion of the first qubit-oscillator unitary, $\hat U_{\text{r}}=\hat U_{\text{m}}^{\dagger}$, which is not viable in our current setup. One way to do this is by having the option to flip the sign of the Hamiltonian. With flux tunability on the qubit frequency~\cite{valadares2024demand}, one might be able to flip the sign of the dispersive coupling $\chi_{\text{qc}}$~\cite{blais2021circuit}, which is the dominant term in the bare Hamiltonian $\hat H_0$, and therefore, allowing the required unitary inversion.




\section{Comparison between OREO and the Hadamard test} 

\begin{figure}
\centering
\includegraphics{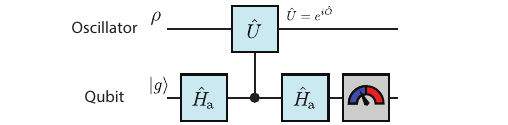}
\caption{Hadamard test. Here, $\hat H_{\text a}$ denotes the Hadamard gate and the control unitary is written as $|g\rangle\langle g|\otimes \mathbb{I} + |e\rangle \langle e |\otimes \hat U$ on the qubit-oscillator's space.}
\label{rev3_figsm_Hadamard}
\end{figure}

Here, we compare our method OREO to the Hadamard test, the sequence of which is given in Fig.~\ref{rev3_figsm_Hadamard}.
The distinctions lie in two aspects: 1. The Hadamard test allows one to get the expectation value of a unitary $\langle \hat U\rangle$, not a Hermitian operator $\langle \hat O\rangle$, which is the goal of our protocol. 2. To get $\langle \hat O\rangle$ using the Hadamard test, one needs much more resources (less efficient) to the point it would be comparable to performing full-state quantum tomography, which is our initial point of comparison. Let us explain in more detail below.

From the Hadamard test, shown in Fig.~\ref{rev3_figsm_Hadamard}, one gets the probability of the ancillary qubit being in the ground state at the end 
\begin{equation}
    p_g=\frac{1+\text{Re}(\langle \hat U \rangle)}{2},
\end{equation}
where $\text{Re}(\langle \hat U \rangle) = (\text{tr}(\rho \hat U)+\text{tr}(\rho \hat U^{\dagger}))/2$.
By doing another setting of the experiment, but now adding a qubit phase rotation to the first Hadamard such that the state of the qubit transforms as $|g\rangle \rightarrow (|g\rangle -i |e\rangle)/\sqrt{2}$, one gets the imaginary part of the expectation value from the qubit measurement: $p_g=(1+\text{Im}(\langle \hat U \rangle))/2$. Combining these outcomes, one can get $\langle U \rangle$.

The ability to obtain $\langle \hat U\rangle$ does not translate to knowledge on $\langle \hat O\rangle$; $\hat U$ and $\hat O$ are different types of operator. For example, one might naively construct the control unitary such that $\hat U=e^{i\hat O}$. However, in general, $\langle \hat U\rangle \ne e^{i\langle \hat O\rangle}$ for arbitrary initial states $\rho$. Therefore, the Hadamard test does not give the same result as OREO, which directly obtains the expectation value of any arbitrary Hermitian operator $\langle \hat O\rangle$.

Now, we would like to discuss two possible approaches we have considered to leverage Hadamard test to obtain $\langle \hat O\rangle$, although \emph{both of which require more resources}.

The first option is, for a given arbitrary $\hat O$, one can express it in terms of generalised Pauli operators, $\hat O=\sum_{i=1}^{D^2} c_i\hat P_i$, where $D$ is the truncation dimension and all $\{P_i\}$ are unitary. To reconstruct an arbitrary Hermitian operator with $D$ dimension, it requires $D^2$ Pauli matrices. One can then run the Hadamard test for $2\times D^2$ different measurement settings to get $\{\langle \hat P_i\rangle\}$, then estimate $\langle \hat O\rangle=\sum_{i=1}^{D^2} c_i\langle \hat P_i \rangle$. In this case, not only we require many settings, but we also need to implement the control unitary such that $\hat U=\hat P_i$. This is resource intensive and it is even worse than the full-state quantum tomography with $D^2-1$ minimal observables that we performed in our lab~\cite{krisnanda2025demonstrating}, which is our initial point of comparison.

The second method involves implementing a control unitary such that $\hat U=\hat O+i\hat K$. Here, both $\hat O$ and $\hat K$ are Hermitian and $\hat O$ is our target observable. The unitary condition $\hat U \hat U^{\dagger}=\hat O^2+ \hat K^2 -i[\hat O,\hat K]=\mathbb{I}$ indicates that $\hat O^2 + \hat K^2=\mathbb{I}$ and $[\hat O,\hat K]=0$ have to be satisfied. We can decompose $\hat O=\sum_k \lambda_k |k\rangle \langle k|$ and choose $\hat K=\sqrt{\mathbb{I}-\hat O^2}=\sum_k \sqrt{1-\lambda_k^2} \:|k\rangle \langle k|$. For $\lambda_k^2\le 1$, a condition imposed by rescaling the given arbitrary observable, the operator $\hat K$ is Hermitian (as required) and one can run the Hadamard test in Fig.~\ref{rev3_figsm_Hadamard} to obtain $p_g=(1+\text{Re}(\langle U\rangle))/2=(1+\text{tr}(\rho \hat O))/2$, from which $\langle \hat O\rangle$ can be extracted. However, this method requires implementing a total unitary $|g\rangle\langle g|\otimes \mathbb{I} + |e\rangle \langle e |\otimes \hat U$ on the qubit-cavity space. This has a couple of serious drawbacks compared to our technique: 1. This method requires universality on the qubit-cavity space, instead of only on the cavity space required by OREO. 2. Although the universality on the qubit-cavity space is known~\cite{eickbusch2022fast}, targeting unitary on a bigger dimension $2\times D$ is much less efficient compared to targeting an operator with $D$ dimension that we do for OREO (explained in detail in Section III. A). Both issues may lead to non-convergence. For example, we attempted to optimize the control pulses to target the qubit-cavity unitary to retrieve the $x$ quadrature, using a truncation dimension $D=10$ on the cavity. Over several independent runs, each lasting approximately one day, none converged. In contrast, using OREO, the optimization takes $\sim 20$ minutes. Therefore, not only is convergence not achieved when optimizing to implement the Hadamard test, but the inefficiency also renders the approach impractical for laboratory implementation, as system parameters drift over time.

In the opposite scenario, with OREO, we can do the Hadamard test, even more efficiently. This follows since the arbitrary unitary can always be decomposed as $\hat U=\hat O_1+i\hat O_2$, where $\hat O1$ and $\hat O_2$ are Hermitian operators. To get $\langle\hat U\rangle$, one needs to run OREO twice (the same as in the Hadamard test to get the real and imaginary parts of $\langle \hat U\rangle$) to get $\langle \hat O_1\rangle$ and $\langle \hat O_2\rangle$. As stated above, OREO requires universality on the oscillator's space only and its optimisation on the truncated space of the oscillator is significantly more efficient.

Thus, while there is indeed a conceptual connection between the Hadamard test and our work, OREO is a much more direct method for extracting the expectation value of an arbitrary operator on the oscillator and significantly less demanding in both experimental and computational resources. 

\section{Simulation of imperfections} 
In this section, we explore the effect of various forms of imperfections on OREO. The system was modeled using a standard master equation simulation using the parameters listed in Table~\ref{table_para} unless otherwise stated.

\subsection{Hamiltonian parameter miscalibration}

\begin{figure}[!h]
\centering
\includegraphics{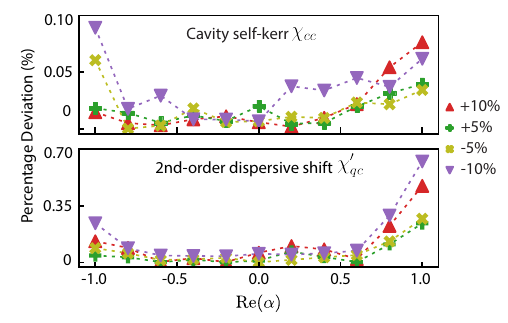}
\caption{Effect of potential differences between the estimated and actual values  of $\chi_{cc}$ and $\chi'_{qc}$  on $\langle x \rangle$ observable measurement}
\label{sfig2}
\end{figure}

Extracting the higher-order Hamiltonian terms ($\chi_{cc}$ and $\chi'_{qc}$) of the system to a high degree of precision is experimentally challenging. As a result, the actual values might deviate from the estimated values by a non-negligible amount. To study the robustness of the protocol to such errors, we simulated the $\langle x \rangle$ measurement using values of $\chi_{cc}$ and $\chi'_{qc}$ that were $ \pm 5{-}10\%$ from the estimated values used to generate the GRAPE pulse. From Fig.~\ref{sfig2}, we can see that the effects of these miscalibrations are small ($< 1\%$).

\subsection{Losses during GRAPE pulses}

\begin{figure}[!htb]
\centering
\includegraphics{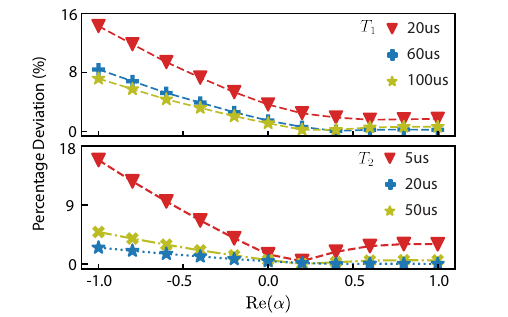}
\caption{Effect of qubit $T_1$ and $T_2$ on $\langle x \rangle$ the values of the phase-space quadrature obtained via OREO.}
\label{sfig3}
\end{figure}

We investigate the impact of transmon decoherence on the performance of OREO. As a reference, we compare simulations of the $\langle x \rangle$ measurement with different $T_1$ and $T_2$ to simulations with a lossless transmon. In Fig.~\ref{sfig3}, we observe significant deviations from the lossless case with typical coherence times in our devices, that is, $T_1\approx 100\,\mu$s and $T_2\approx40\text{-}50\,\mu$s. Thus, the performance of OREO will directly benefit from the mainstream pursuits of the cQED community to continually improve the qubit coherence properties. Furthermore, we also observe that the errors become more pronounced when the coherence state is displaced further out in the phase space. This is consistent with our analysis that the choice of truncation dimension is a crucial factor in OREO to ensure its reliability.

\subsection{Truncation dimension}
In our experiments, we use $D=6$ as the truncation dimension when carrying out the optimization in OREO, see Eq.~(\ref{EQ_Oeffsigmaz}). For states that are further away in phase space, e.g., the coherent state $|\alpha=-1+i\rangle$ (the top left corner of Fig.~2a in the main text, $D=6$ contains $98\%$ of the total population, and therefore, introduces error to the expectation value of $\langle x \rangle$.

\begin{figure}[!h]
\centering
\includegraphics{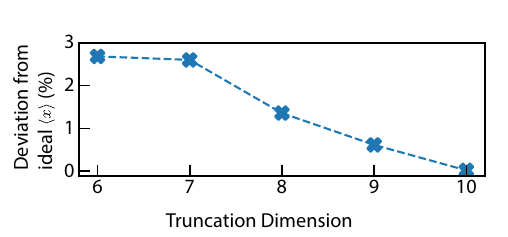}
\caption{Effect of truncation dimension on $\langle x \rangle$ observable measurement for the coherent state $|\alpha=-1+i\rangle$.}
\label{sfigtrun}
\end{figure}

To isolate the error caused by the truncation dimension, we simulate the value $\langle x\rangle$ for the coherent state $|\alpha=-1+i\rangle$ without decoherence, for different truncation dimensions used in optimization for $\hat U_{\text{m}}$. Fig.~\ref{sfigtrun} shows the error in $\langle x \rangle$ drops as we increase the truncation dimension. 

\blue{We note that the optimization performed in OREO does not constitute a substantial overhead, as it is carried out only \emph{once} prior to the experiment, after which it is applicable to any initial state within the truncation dimension $D$. Moreover, we show that this optimization is not a limiting factor: it converges quickly to a very small cost function value, $\mathcal{C} = 10^{-5}$. Table~\ref{tab:time_D1} reports the average optimization time (over five trials) required to reach this cost for different truncation dimensions $D$, using a computer equipped with an AMD Ryzen~9~7950X 16-core processor and 64~GB RAM.}  

\begin{table}[h]
\centering
\begin{tabular}{|c|c|}
\hline
$D$ & Time (mins) \\ \hline \hline
6 & 6 \\ \hline 
7 & 7 \\ \hline 
8 & 10 \\ \hline
9 & 13 \\ \hline 
10 & 19 \\ \hline
\end{tabular}
\caption{\blue{Average optimization time required to reach a cost function value of $\mathcal{C} = 10^{-5}$ for different truncation dimensions $D$.}}
\label{tab:time_D1}
\end{table}

\blue{From our perspective, the main limitation at large truncation dimensions is the signal-to-noise ratio. In particular, for observables with high dimension whose eigenvalues extend far beyond the range $[-1,1]$ (the range of the qubit measurement $\langle \sigma_z \rangle$), such as quadratures, the scaling factor $f$ becomes large. As a result, the noise in $\langle \sigma_z \rangle$ is amplified proportionally to $f$. This, however, is a fundamental limitation of any method that seeks to retrieve high-dimensional observables from a qubit measurement, rather than a limitation specific to our approach.}


\section{Calibration and correction for readout assignment error}

As the value of the observable is directly mapped to the qubit state, it is important to account for qubit state assignment (readout) errors. To characterize the readout error, we first prepare the qubit in the $|g\rangle$ ($|e\rangle$) state and measure the $|e\rangle$ population to obtain $p_{ge}$ ($p_{ee}$), which was found to be 0.02 (0.96). This value is then used to offset (scale) the measured excited population using the formula:
\begin{equation}
    p_e^{\text{corrected}} = \frac{p_e^{\text{measured}} - p_{ge}}{p_{ee}}
\end{equation}

\section{non-Gaussianity inference}

\subsection{Hierarchy and threshold for Fock states and their superpositions}
Quantum non-Gaussian states are those whose Wigner functions deviate from classical mixture of Gaussian states~\cite{filip2011detecting} and, eventually, can exhibit features such as negative values in their Wigner function. These states are resources for universal quantum computing~\cite{braunstein2005quantum}, quantum error correction~\cite{ofek2016extending, sivak2023real, ni2023beating}, and quantum metrology~\cite{pan2025realization, deng2024quantum}. Some non-Gaussian states have features that are different than a classical mixture of others, and therefore, it is important to formulate and test the non-Gaussianity hierarchy.

In this work, we explore a hierarchy of $n$-photon quantum non-Gaussianity, which asserts that a state with a non-Gaussianity rank of $n$ is incompatible with any mixture of Fock state superpositions up to Fock $|n-1\rangle$ with the addition of any application of Gaussian operation~\cite{lachman_faithful_2019}. Mathematically, a pure state $|\psi \rangle$ is said to have rank $n$ if it cannot be written as 
\begin{equation}
    |\psi \rangle \ne \hat G |\tilde 
    \psi_{n-1}\rangle,\label{EQ_GMix}
\end{equation}
where $\hat G$ is a Gaussian operation, which can be implemented by a displacement and/or squeezing operation, and $|\tilde \psi_{n-1}\rangle=\sum_{k=0}^{n-1} c_k |k\rangle$ is arbitrary superposition of Fock states up to Fock $|n-1\rangle$.



\begin{figure}[!htb]
\centering
\includegraphics{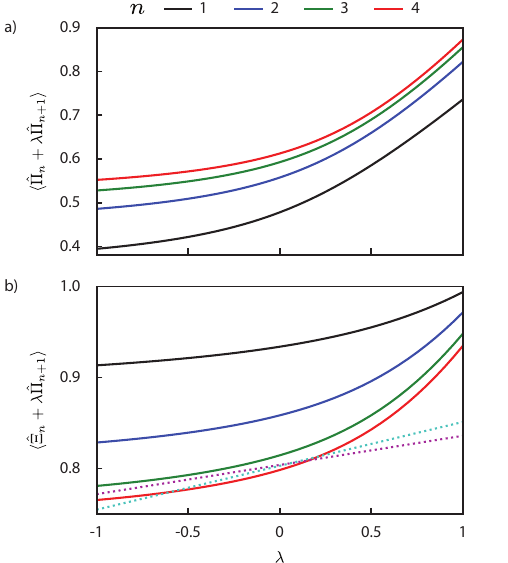}
\caption{Thresholds for the observable $\hat\Pi_n+\lambda \hat\Pi_{n+1}$ in a) and $\hat \Xi_n+\lambda \hat \Pi_{n+1}$ in b) with a given parameter $\lambda$. The curves depict the thresholds with the rank $n=1$ (black),
$n=2$ (blue), $n=3$ (green) and $n=4$ (red). An outcome of a measurement exceeding the corresponding threshold certifies $n$-photon quantum non-Gaussianity. Dotted magenta and cyan lines show computed observable for the reconstructed binomial state $(|0\rangle+|4\rangle)/\sqrt{2}$ obtained from projecting a vacuum (Fig.~4b(ii)) and thermal state (Fig.~4c(ii)), respectively. Both cases show values slightly above the threshold for rank $n=4$ (red curve). This is due to the use of weak readout drive for the qubit measurement, similar to the case of weak dephasing drive in Fig.~3d for the state $(|0\rangle+|3\rangle)/\sqrt{2}$.}
\label{sfig_threshold}
\end{figure}

To probe the rank experimentally, we measure a single observable $\hat O$. For an arbitraty state $\rho$, if the expectation value of the observable $\langle \hat O\rangle_{\rho}>F_n$, where $F_n$ is the rank-$n$ threshold, one concludes that the state has rank $n$. The threshold $F_n$ is computed by optimizing the observable $\langle \hat O \rangle$ given the state $\hat G |\tilde \psi_{n-1}\rangle$ over the parameters of the Gaussian operation and superposition weights $\{c_k\}$ as in Ref.~\cite{lachman_faithful_2019}. While optimizing over the superposition weights $\{c_k\}$ corresponds to an eigenvalue problem~\cite{fiurasek_efficient_2022}, the optimum over the Gaussian operation is reached only numerically \cite{lachman_faithful_2019}. We chose first an observable $\hat O_n=\hat \Pi_n+\lambda \hat \Pi_{n+1}$, where $\hat \Pi_n\equiv|n\rangle \langle n|$ and $\lambda$ is a given real parameter, to probe rank $n$ non-Gaussianity of Fock state $|n\rangle$. This choice is motivated by the incorporation of successful probability $ \langle \hat\Pi_n \rangle$ and a leakage error $ \langle \hat\Pi_{n+1} \rangle$ into a higher Fock state. 
\blue{Moreover, the free parameter $\lambda$ does not represent a metric nor quantity certifying a degree of quantum non-Gaussianity. It instead serves as a tool to extend the class of measurements suitable for certification and enlarge the set of states that can be detected as quantum non-Gaussian.  It is because surpassing a corresponding threshold for $\langle \hat \Pi_n+\lambda \hat \Pi_{n+1}\rangle$ with any value of $\lambda$ is sufficient for the hierarchical certification, we can optimize the parameter $\lambda$ for the given data to overcome the threshold. Fig.~3 (main text) presents thresholds for various $\lambda$ and compare them with the data. The threshold grows monotonically with $\lambda$, reaches always a positive value even for very small negative $\lambda$ and converges to zero in the limit $\lambda\ll-1$. This implies that states with $\langle \hat \Pi_{n+1}\rangle=0$ and $\langle \hat \Pi_{n}\rangle>0$ exhibit the quantum non-Gaussianity when choosing measurement with sufficiently small $\lambda$.}

Similarly, for the superposition of Fock state $(|0\rangle + |n\rangle)/\sqrt{2}$, we chose the observable $\hat O=\hat \Xi_n+\lambda \hat \Pi_{n+1}$, where $\hat \Xi_n\equiv|0\rangle \langle n|+|n\rangle \langle 0|$, capturing the amplitude of the off-diagonal elements~\cite{kovalenko2025quantum, asenbeck2025hierarchical, lachman2025hierarchies}. The numerically optimized thresholds for $\hat \Pi_n+\lambda \hat \Pi_{n+1}$ and $\hat \Xi_n+\lambda \hat \Pi_{n+1}$ are shown in Fig.~\ref{sfig_threshold}a and b, respectively. Figs.~3b and 3d in the main text depict the thresholds for $\hat \Pi_3+\lambda \hat \Pi_{4}$ and $\hat \Xi_3+\lambda \hat \Xi_{4}$ as the lower boundary of the blue shaded region.

\subsection{Photon loss}
\begin{figure*}[t]
\centering
\includegraphics{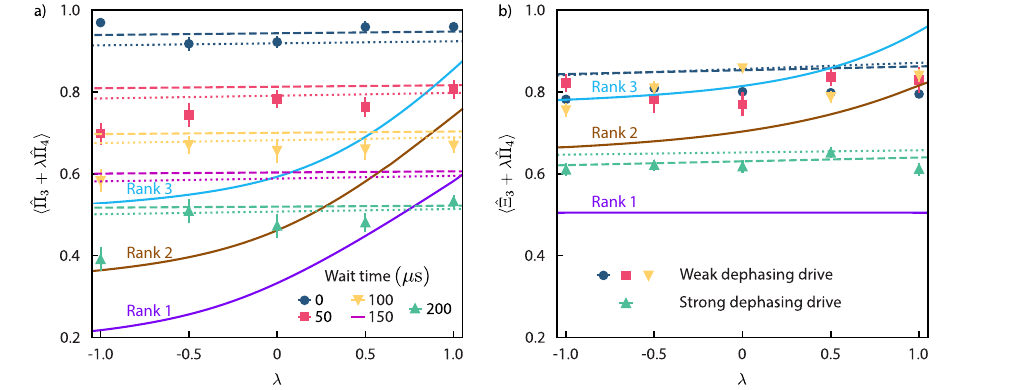}
\caption{\blue{(a) Impact of photon loss on the non-Gaussianity rank of the state $|3\rangle$ and (b) impact of dephasing on the non-Gaussianity coherence rank of the state $(|0\rangle+|3\rangle)/\sqrt{2}$. In both panels the dotted lines represent simulations of the expectation value of the corresponding observable for the reconstructed density matrix under various loss cases, while the dashed lines represent full simulations with decoherence. Markers correspond to the experimental data presented in Fig.~3 of the main text. The purple, brown, and cyan solid curves respectively correspond to the first, second, and third order non-Gaussianity threshold above which one can infer the corresponding rank. The magenta dotted and dashed lines in panel (a) indicate that, at a wait time of $150$~$\mu$s, the expectation value for $\lambda = 0$ reaches the non-Gaussianity threshold (cyan solid curve).}}
\label{sfig7}
\end{figure*}

It is evident from Fig.~3b in the main text that photon loss degrades the quality of the prepared Fock state $|3\rangle$. We also observe that having flexibility in the free parameter $\lambda$ enables us to infer a non-Gaussianity rank of $3$ for wait times up to $100$~$\mu$s. With the full density matrix reconstructed via Wigner tomography, we simulate the effect of photon loss on the oscillator's density matrix and compute the expectation value of the observable $\langle \hat{\Pi}_3 + \lambda \hat{\Pi}_4 \rangle$. These simulations are shown as dotted lines in Fig.~\ref{sfig7}a, demonstrating good agreement with the experimental data (represented by the corresponding markers). \blue{Moreover, we also performed full simulations with decoherence, which are shown as dashed lines in Fig.~\ref{sfig7}a. }

\blue{Notably, at a wait time of $150$~$\mu$s, indicated by the magenta lines (both dotted and dashed), the expectation values for $\lambda = 0$ lie close to the non-Gaussianity threshold (solid cyan curve), where inference of rank $3$ from experimental data will likely be inconclusive. However, the magenta lines remains above the threshold for $\lambda < 0$, suggesting that it is still possible to infer a rank of $3$. This highlights the significance of considering the $\lambda \neq 0$ scenario, which is non-trivial, as it cannot be realized by a single photon number measurement native to the cQED platform. Moreover, we note that although at a wait time of $200$~$\mu$s, we cannot infer rank $3$ from experimental data (green triangles), they are still above the rank $2$ threshold (brown curve), and therefore they satisfy rank $2$ non-Gaussianity.}

In Fig.~\ref{sfig7}, we present the thresholds for rank $1$, $2$, and $3$ according to the value of the corresponding observable ($\langle\hat \Pi_3+\lambda \hat\Pi_4\rangle$ in panel (a) and $\langle\hat \Xi_3+\lambda \hat\Pi_4\rangle$ in panel (b)).
The ordered threshold for the corresponding rank $n$ covers the states $\hat G |\tilde{\psi}_{n-1}\rangle$ in Eq.~(\ref{EQ_GMix}).
Note that this is different compared to Fig.~\ref{sfig_threshold} where a threshold with the highest rank is always plotted for a given observable. 

\subsection{OREO vs full-state tomography}

\blue{OREO achieves the estimate $\mathrm{tr}(\rho \hat O)$ of an arbitrary oscillator state $\rho$ for a given observable $\hat O$ in a single-observable measurement. Since analytical gate sequences beyond those used to map excitation number $\mathrm{tr}(\rho |n\rangle \langle n|)$~\cite{schuster2007resolving}, parity $\mathrm{tr}(\rho e^{i\pi\hat a^{\dagger}\hat a})$~\cite{sun2014tracking}, and the real and imaginary parts of the characteristic function $\mathrm{tr}(\rho \hat D(\alpha))$~\cite{campagne2020quantum} do not exist, the conventional way to obtain such estimates is through full-state tomography. In this approach, one first reconstructs the density matrix $\rho$ of the oscillator state and then computes $\mathrm{tr}(\rho \hat O)$. This procedure provides the natural benchmark against which we compare our OREO method.
}

\blue{In particular, we employed a standard grid-based Wigner tomography. This is done as follows. We employed the analytical gate sequence that maps the parity of the oscillator's state~\cite{sun2014tracking} displaced to different regions of phase space. In other words, we measured $\mathrm{tr}\!\left(\hat D(-\alpha)\,\rho\, \hat D(\alpha)\, e^{i\pi \hat a^{\dagger}\hat a}\right)$,  
for a set of displacement points $\{\alpha\}$ arranged on a $21\times 21$ grid. At each point, we performed $1000$ repetitions to obtain the statistics necessary for evaluating the expectation value. Since the parity mapping protocol is known to suffer from both coherent and incoherent errors~\cite{krisnanda2025demonstrating}, we additionally implemented the standard corrected-parity procedure for every displacement point, effectively doubling the number of observables. In total, this amounts to $882$ distinct observables, each repeated $1000$ times. From the measurement results, linear inversion and Bayesian inference can then be used to retrieve the reconstructed density matrix~\cite{krisnanda2025demonstrating}.}

\blue{By contrast, OREO requires only a single-observable measurement, repeated $1000$ times, to extract $\mathrm{tr}(\hat O\rho)$. This reduces the acquisition time by a factor of $882$, i.e., nearly \emph{three orders of magnitude}, while giving the same results, see for example Fig.~3 in the main text. 
}






\section{Impact of induced dephasing on superpositions of Fock States}
A very device-specific imperfection in this experiment is the relatively large cross-Kerr interaction term $-\chi_{cr} \hat c^{\dagger}\hat c \hat r^{\dagger}\hat r \nonumber$ between the cavity and the readout resonator with $\chi_{cr}\approx 2\,$kHz, which causes unwanted dephasing and rotation in phase space of the cavity state when the readout resonator is populated. Thus, we can induce different rates of dephasing on the cavity by varying amplitude and duration of a drive applied to the readout resonator. In this investigation, we experimented with two such pulses of the same duration of $1.68$~$\mu$s but different amplitude. 

\begin{figure}[!htb]
\centering
\includegraphics{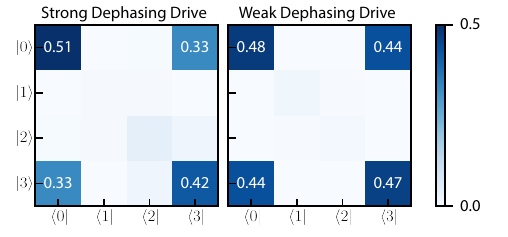}
\caption{Real components of the reconstructed density matrix of the $|0\rangle + |3\rangle$ state under strong (weak) readout. The strong readout pulse results in a larger dephasing effect which can be observed by the higher reduction of the off-diagonal $|0\rangle \langle  3| $ and $ |3\rangle \langle  0| $ components. The amplitude of the weak readout pulse was $\approx 42\%$ of the strong readout pulse.}
\label{sfig4}
\end{figure}


As shown in Fig.~\ref{sfig4}, the resulting dephasing reflected by the reduction in the off-diagonal elements is more pronounced when we use a pulse with higher amplitude, i.e. more photons are introduced in the readout resonator. Furthermore, knowing the reconstructed density matrix of the dephased state, we can estimate the corresponding coherence time of the oscillator due to the presence of the resonator photons. Our simulations show that the coherence times in the presence of these dephasing drives are approximately $T_{\phi,c}^{\text{strong}}\approx42$~$\mu$s and $T_{\phi,c}^{\text{weak}}\approx263$~$\mu$s for the strong and weak pulse, respectively.

We note that for weaker pulse amplitude, the resulting distributions the IQ plane of the readout resonator, corresponding to the qubit states $|g\rangle$ and $|e\rangle$, overlap. In this situation, we can still extract reliable information about the qubit state by adjusting the threshold for single-shot state discrimination to increase confidence in identifying the qubit in the ground state at the cost of discarding more shots. Specifically, this means that only approximately $15\%$ of the data, corresponding to cases where the qubit is actually in $|g\rangle$, is selected, while the remaining $\approx 85\%$, including the overlapping region where $|g\rangle$ and $|e\rangle$ cannot be distinguished, is discarded.

To verify that the reduction in $\langle \hat{\Xi}_3 + \lambda \hat{\Pi}_4\rangle$ is indeed a result of the induced dephasing effect, we performed the same experiments as in Fig.~3d of the main text, now using a weaker pulse for the qubit measurement following state preparation, and prior to the application of the mapping unitary $\hat{U}_{\text{m}}$. 

All results consistently show that the use of a weak readout pulse, which improves the coherence time $T_{\phi,c}$ by a factor of 6, increases the expectation value of the observable, see Fig.~3d in the main text. However, the results remain close to the rank-3 threshold, with only a few points exceeding it, depending on the choice of pulses used for implementing $\hat{U}_{\text{m}}$. Designing a device with reduced cross-Kerr interactions, or using an even weaker readout pulse (at the expense of discarding more data), would further improve the coherence of the superposition state and enable a conclusive inference of the non-Gaussianity rank.

\blue{With the current experimental data, we can confidently infer that the prepared state with weak dephasing drive has non-Gaussianity coherence rank $2$, whereas it is rank $1$ for the case of strong dephasing drive, see Fig.~\ref{sfig7}b.}


\section{Preparation of thermal state}

The thermal state was prepared by first initializing the transmon in $|g \rangle$ and the cavity in $| \alpha = 0.5 \rangle$. To induce decoherence in the cavity state, a $\pi/2$-pulse is applied to the transmon to put it into a superposition $(|g\rangle + |e\rangle)/\sqrt{2}$. The system is then allowed to evolve for $\tau=452$ns. The dispersive interaction between the cavity and the transmon causes the system to form an entangled state, which quickly degrades into a statistical mixture due to transmon decoherence~\cite{valadares2024demand}. 

To amplify this effect, the readout resonator is populated via a strong drive, causing accelerated transmon decoherence due to photon shot noise~\cite{sears2012photon}. This process is repeated 40 times to fully scramble the oscillator state and increase its effective temperature. Finally, the transmon is projected to the $|g \rangle$ state.
\blue{This procedure will result in a completely dephased state that is theoretically closest to a thermal state with mean photon number $n_{\text{th}} = 0.24$.}

\blue{We note that the completely dephased state has a theoretical fidelity of $0.78$ with the vacuum state. In experiments, this is confirmed by the fidelity of the reconstructed density matrix for the thermal state and vacuum, which is $0.78(2)$, whereas its fidelity with thermal state with $n_{\text{th}}=0.24$ is $0.97(2).$}

\section{Sequential projections}

The projective measurement experiment in the main text can be further extended by performing two projective measurements in a row to increase the yield of successful projections to the $|\psi\rangle = (|0\rangle+|4\rangle)/\sqrt{2}$ state. Although a failed projection leads to a state with a very small $|\psi\rangle$ component, a displacement pulse $\hat D(\alpha = -1)$ played to the cavity can take it to a non-orthogonal state, increasing the probability of success in the next iteration.

\begin{figure}[!htb]
\centering
\includegraphics{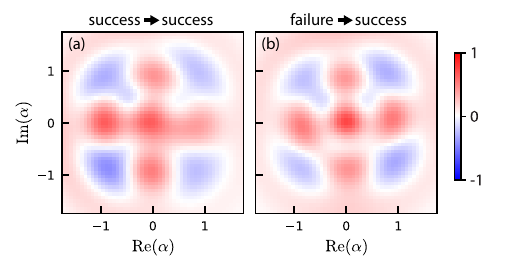}
\caption{Reconstructed Wigner functions of the final cavity states after two rounds of projective measurements for the cases where (a) the qubit is measured in $|g \rangle$ during both iterations and (b) the qubit is measured in $|e \rangle$ after the first iteration and in $|g \rangle$ after the second one.}
\label{sfig5}
\end{figure}

Fig.~\ref{sfig5} shows the reconstructed Wigner functions of the final state at the end of 2 iterations. The fidelity of the final state after 2 successful projections to the ideal binomial state was about 0.77 while the fidelity of the state with 1 failed and 1 successful projection was about 0.70. This reduction in fidelity is consistent with the limit imposed by transmon decoherence during the repeated applications of the numerically optimized pulses. 

\section{How to use OREO in your experiment}

All code used in our work is available on GitHub: \url{https://github.com/tkrisnanda/OREO}. Below, we briefly outline the steps to implement OREO in your experiment:
\begin{enumerate}
    \item \textbf{Measure system parameters.} Experimentally determine the relevant Hamiltonian parameters for the qubit-oscillator subsystem, which for our setup correspond to the values in Table~\ref{table_para}. Specifically, we measure
    \begin{enumerate}
        \item Transmon frequency $\omega_q$ and self-Kerr $\chi_{qq}$ through spectroscopy, fine-tuned through Ramsey interferometry.
        \item Cavity frequency $\omega_c$ through spectroscopy.
        \item Cavity self-Kerr $\chi_{cc}$ first guess from simulation. The guess is fine-tuned by tracking the time evolution of the quadrature $\langle x\rangle$ for coherent states of different amplitudes, which gives the frequency deviation with increasing cavity population.
        \item Transmon-cavity coupling parameters $\chi_{qc}$ and $\chi_{qc}'$ by measuring the transmon frequency shift when the cavity is prepared in $|1\rangle$ and $|2\rangle$. \\
    \end{enumerate}
    \item \textbf{Calibrate gates to implement universal operations on the oscillator.} The amplitudes of numerically optimized pulses are calibrated by testing single-qubit and single-cavity operations. The calibrated numerical optimization is confirmed by creating more complex states such as cat and Fock states. 
    \item \textbf{Choose a truncation dimension $D$ for the observable $\hat{\mathcal{O}}$.} This dimension should be large enough to capture at least 99\% of the total population of the oscillator states relevant to the measurement. In our implementation, we chose $D = 6$.
    \item \textbf{Run the optimization to determine the time-dependent oscillator and qubit drives that implement the mapping unitary $\hat{U}_{\text{m}}$.} Inputs to the optimization include: the target observable (a scaled version $\hat{O}$), the relevant Hamiltonian parameters (listed in Table~\ref{table_para}), and the desired operation duration. The optimization is performed without including decoherence effects. See our GitHub repository for details.
    \item \textbf{Follow the OREO procedure in your experiment:} prepare an initial state of your choice, apply the optimized drives to implement $\hat{U}_{\text{m}}$, and perform single-shot qubit measurements.
    \item \textbf{Post-process the results to obtain $\langle \hat{\mathcal{O}} \rangle$.} Perform corrections to account for readout error. Then, retrieve the expectation value of the original observable: $\langle \hat{\mathcal{O}} \rangle = f \times \langle \hat{\sigma}_z \rangle$.
\end{enumerate}
To implement the oscillator state projection, an additional optimization is required to determine the time-dependent oscillator and qubit drives for the unitary $\hat{U}_{\text{r}}$. The input includes: the operator $\langle g|\hat{U}_{\text{m}}^{\dagger}|g\rangle$, computed from the previously obtained drives for $\hat{U}_{\text{m}}$; the Hamiltonian parameters from Table~\ref{table_para}; and the operation duration. Carry out the experimental sequence as shown in Fig.~4a of the main text.

\end{document}